\documentclass[a4paper,referee]{raa}            

\usepackage{graphicx,times}             
\usepackage{natbib}
\usepackage{amssymb,amsmath}
\usepackage{overpic}
\usepackage{orcidlink}
\bibpunct{(}{)}{;}{a}{}{,}
\begin{document}

  \title{Low surface brightness galaxies from BASS+MzLS with Machine Learning}
\volnopage{ {\bf 20XX} Vol.\ {\bf X} No. {\bf XX}, 000--000}
  \setcounter{page}{1}

\author{Peng-liang Du\inst{1,2}~\orcidlink{0009-0008-0836-6732}, 
Wei Du\inst{1}~\orcidlink{0000-0003-4546-8216}, 
Bing-qing Zhang\inst{1}~\orcidlink{0000-0002-6659-1152}, 
Zhen-ping Yi\inst{3}~\orcidlink{0000-0001-8590-4110}, 
Min He\inst{1}~\orcidlink{0000-0001-6139-7660}, 
Hong Wu\inst{1,2}~\orcidlink{0000-0002-4333-3994}
}
\institute{ Key Laboratory of Optical Astronomy, National Astronomical Observatories, Chinese Academy of Sciences, Beijing,100101,China; {\it wdu@nao.cas.cn}\\
        \and
          School of Astronomy and Space Science, University of Chinese Academy of Sciences, 19A Yuquan Road, Shijingshan District, Beijing, 100049, China\\
	\and
      School of Mechanical, Electrical and Information Engineering, Shandong University, 180 Wenhua Xilu, Weihai, 264209, Shandong, People’s Republic of China\\
\vs \no
   {\small Received 20XX Month Day; accepted 20XX Month Day}
}



\abstract{From $\sim$ 5000 deg$^{2}$ of the combination of the Beijing-Arizona Sky Survey (BASS) and Mayall $z$-band Legacy Survey (MzLS) which is also the northern sky region of the Dark Energy Spectroscopic Instrument (DESI) Legacy Imaging Surveys, we selected a sample of 31,825 candidates of low surface brightness galaxies (LSBGs) with the mean effective surface brightness 24.2 $< \bar{\mu}_{\rm eff,g} <$ 28.8 mag arcsec$^{\rm -2}$ and the half-light radius 2.5$^{\prime\prime}$ $< r_{\rm eff} <$ 20$^{\prime\prime}$ based on the released photometric catalogue and the machine learning model. The distribution of the LSBGs is of bimodality in the $g$ - $r$ color, indicating the two distinct populations of the blue ($g$ - $r <$ 0.60) and the red ($g$ - $r >$ 0.60) LSBGs. The blue LSBGs appear spiral, disk or irregular while the red LSBGs are spheroidal or ellipitcal and spatially clustered. This trend shows that the color has a strong correlation with galaxy morphology for LSBGs. In the spatial distribution, the blue LSBGs are more uniformly distributed while the red ones are highly clustered, indicating that red LSBGs preferentially populated denser environment than the blue LSBGs. Besides, both populations have consistent distribution of ellipticity (median $\epsilon\sim$ 0.3), half-light radius (median $r_{\rm eff} \sim$ 4$^{\prime\prime}$), and S$\rm \Acute{e}$rsic index (median $n$ = 1), implying the dominance of the full sample by the round and disk galaxies. This sample has definitely extended the studies of LSBGs to a regime of lower surface brightness, fainter magnitude, and broader other properties than the previously SDSS-based samples.
}

   \authorrunning{P.-L. Du et al. }            
   \titlerunning{LSBGs from BASS+MzLS}  
   \maketitle

\section{INTRODUCTION} \label{sec:intro}
Low surface brightness galaxies (LSBGs) are traditionally defined as galaxies with the $B$-band central surface brightnesses ($\mu_{\rm 0}$) fainter than a threshold value within 21.65 - 23.0 mag arcsec$ ^{-2} $\citep{1970ApJ...160..811F,1997ARA&A..35..267I,1997AJ....113.1212O,2008MNRAS.391..986Z,2015AJ....149..199D}. In addition, the $\mu_{\rm 0}$ in some other optical or near infrared bands such as the $r$ \citep{1996ApJS..103..363C}, $R$ \citep{2006A&A...459..679A}, and $K_{\rm S} $ bands \citep{2003A&A...405...99M} have been adopted to distinguish between LSBGs and high surface brightness galaxies (HSBGs) as well.  Besides the $\mu_{\rm 0}$, the mean surface brightness within effective radius ($\bar{\mu}_{\rm eff}$) has also been utilized to define LSBGs, for example,  the criterion of the $g$-band $\bar{\mu}_{\rm eff} > $ 24.2 - 24.3 mag arcsec$ ^{-2}$ was once used to select LSBGs in \citet{2018ApJ...857..104G,2021ApJS..252...18T}, allowing for the retention of nucleated galaxies in the sample. \par

LSBGs are characterized by their diffuse, extended, low-density stellar discs and most of them are blue in color \citep{1996MNRAS.283...18D,2001AJ....122.2318B,2004A&A...428..823O,2006A&A...458..341T,2009A&A...505..483V,2024RAA....24a5018Z}. In morphology, they are disk-like or irregular \citep{1996ApJ...469L..89D,1997MNRAS.290..533D,2001AJ....122.2396D}. Compared to HSBGs, LSBGs have different properties, including low star formation rates \citep{1993AJ....106..548V,1997AJ....114.2497V,2000A&A...357..397V,2009ApJ...696.1834W,2011AdAst2011E..12S,2011ApJ...728...74G,2018ApJS..235...18L,2019ApJS..242...11L,2022ApJ...940L..37G}, low metallicities \citep{1998A&A...335..421D,1998A&A...336...49D,2004MNRAS.355..887K,2017ApJ...837..152D}, high gas fractions \citep{2014ApJ...793...40H,2015AJ....149..199D,2020ApJS..248...33H}, low dust content \citep{2001A&A...365....1M,2007ApJ...663..908R,2007ApJ...663..895H}, and low AGN fraction \citep{2011ApJ...728...74G}, which indicate that LSBGs are different in star formation and evolutionary history from HSBGs. Therefore, it is vital to study LSBGs to complete the current paradigm of galaxy formation and evolution. Moreover, given that LSBGs contributing approximately 20$\%$ \citep{2004MNRAS.355.1303M} to the dynamical mass of the galaxies in the universe and $\sim$ 30$\%$ - 60$\%$ \citep{1995AJ....110..573M,1996MNRAS.280..337M,1997PASP..109..745B,2000AJ....119..136O,2006A&A...458..341T,2007A&A...471..787H,2019MNRAS.485..796M} to the number density of galaxies in the local universe, LSBGs play a significant role in understanding the universe.\par

In the past, researches on LSBGs were primarily concentrated in smaller regions such as massive galaxy clusters \citep{2005MNRAS.357..819S,2015ApJ...798L..45V,2017A&A...608A.142V}, satellites of nearby galaxies \citep{2013ApJ...776...80M,2018ApJ...868...96C}, and other nearby clusters. However, with the advancement of modern observational technology and the emergence of larger, more sensitive telescopes, it has become possible to untargeted search for LSBGs using the deep and wide-field imaging surveys. In the recent decades, wide-field galaxy surveys have revealed a large number of LSBGs. For example, the Sloan Digital Sky Survey \citep[SDSS;][]{2000AJ....120.1579Y} DR4 had established a population of 12,282 face-on LSBGs by \cite{2008MNRAS.391..986Z}. \cite{2018ApJ...857..104G} discovered 781 LSBGs with untargeted search in the Hyper Suprime-Cam Subaru Strategic Program\citep[HSC-SSP;][]{2018PASJ...70S...4A}. Recently, \cite{2021ApJS..252...18T} constructed a large sample of 23,790 LSBGs based on the first three years of data from the Dark Energy Survey \citep[DES;][]{2005astro.ph.10346T}. In addition, the imaging survey of SDSS DR7 and the 40$\%$ Arecibo Legacy Fast ALFA (ALFALFA) Survey \citep{2007NCimB.122.1097G} have been combined to search for samples with low optical surface brightnesses and abundant neutral hydrogen gas \citep{2015AJ....149..199D,2020ApJS..248...33H}. More recently, the candidates of the ultra-diffuse galaxies (UDGs), a subset of LSBG with $g$-band ${\mu}_{\rm 0} \gtrsim 24$ mag arcsec$^{\rm -2}$ and the effective radii $r_{\rm eff} \gtrsim 1.5$ kpc \citep{2015ApJ...798L..45V}, are selected by \citet{2022ApJS..261...11Z,2023ApJS..267...27Z} from the Dark Energy Spectroscopic Instrument (DESI) Legacy Imaging Surveys \citep[hereafter referred to as the Legacy Surveys;][]{2019AJ....157..168D}. \par

In recent years, the advent of more and more deep and wide imaging surveys brought unprecedented opportunities to detect numerous LSBGs with much fainter surface brightness than before. With these samples of more LSBGs with much lower surface brightnesses from the images at previously unreachable depth, the existing LSBG samples that are dominated by brighter LSBGs($<$ 24.0 mag arcsec$^{\rm -2}$) would be highly completed by fainter LSBGs that have much lower surface brightnesses, which would be definitely useful to refine or complete the extant conclusions that are biased towards the LSBGs with brighter surface brightnesses, and provide new constraints on galaxy formation theory and the cosmological models. Thanks to the increasingly widespread application of the computer techniques in the field of astronomy \citep{2020MNRAS.493.4209C}, it is possible to expedite the search for LSBGs amidst the continuously increasing astronomical data with the help of the available computer techniques, such as machine learning. For example, \citet{2021ApJS..252...18T} searched for LSBGs from the data of the first three years of DES observing (DES Y3), utilizing the machine learning techniques. In other work, the deep learning techniques are used to identify LSBGs from the digital sky survey images \citep{2019ApJS..240....1Z,2021A&C....3500469T,2022MNRAS.513.3972Y}. In this paper we are inspired to obtain a catalogue of LSBG candidates from the data from DR9 of the northern portion of the Legacy Surveys in virtue of the machine learning technique. \par

In this paper, we briefly describe the Legacy Surveys and the initial data in Section~\ref{THE DATA}, and described the initial data and the selection of the sample of the LSBG candidates by using the machine learning in Section~\ref{LSBG CATALOG}. We studied the properties of the sample of the LSBG candidates in  Section~\ref{LSBG PROPERTIES}, such as the color distribution, spatial distribution, and other properties. Finally, we compare the LSBG candidates with the LSBG samples from several previous publications in Section~\ref{sect:discussion} and make a summary in Section~\ref{sect:conclusion}.

\section{DATA}\label{THE DATA}
 The DESI Legacy Surveys conducted observations of $\sim$ 14,000 deg$^{2}$ of extragalactic sky in three optical bands ($g$, $r$, $z$). The 5$\sigma$ point source depths for the Data Release 9 (DR9) of the Legacy Surveys are about $g$ = 24.7, $r$ = 23.9, and $z$ = 23.0 AB mag, apparently deeper than those for the SDSS images which are $g$ = 23.13, $r$ = 22.70, and $z$ = 20.71, respectively. Therefore, the data from the Legacy Surveys are expected to embrace numerous galaxies with much lower surface brightness than the SDSS data.  
 
 Additionally, the Legacy Surveys are composed of three imaging projects of the Beijing-Arizona Sky Survey \citep[BASS;][]{2017PASP..129f4101Z}, the Mayall $z$-band Legacy Survey (MzLS), and the Dark Energy Camera Legacy Survey (DECaLS). Specifically, the BASS has surveyed an area of $\sim$ 5500 deg$^{2}$ which is dominated by the region of the sky at Dec $\geq +32^{\circ}$ (with only $\sim$ 4$\%$ located at Dec $\textless +32^{\circ}$) in the optical $g$ and $r$ bands using the Bok 2.3m telescope at Kitt Peak. The MzLS has observed nearly the same sky region as the BASS at Dec $\geq +32^{\circ}$ but in the $z$-band, which well provides a complementary band to extend the band coverage of the BASS. Hereafter, we refer to the survey of both BASS and MzLS at Dec $\geq +32^{\circ}$ as the BASS+MzLS, and intend to select the LSBG candidates from the data of BASS+MzLS.


\section{LSBG CATALOG}
\label{LSBG CATALOG}

In this section, we elaborate the procedures that we followed to select the LSBG candidates from the BASS+MzLS, based on the combination of the publicly photometric catalogue produced by the Tractor software \citep{2016ascl.soft04008L} for DR9 and the machine learning technique.

\subsection{Initial sample selection}\label{Initial sample selection}
The Tractor catalogue of DR9 for the BASS+MzLS provides valuable properties for the total sample of 364,277,779 extracted sources, including astrometry, photometry, and geometry. Based on some crucial properties in this catalogue, we select the LSBG candiates according to the following procedures step by step.\par

First of all, most LSBGs are acknowledged to be dominated by an extended disk, so we remove sources with morphological types (\emph{type}) of PSF, DUP or DEV from the total sample. By this criterion, the point sources, coincident Gaia sources, or elliptical galaxies are excluded and 187,492,198 sources ($\sim51.470\%$) of the total sample are retained.

Secondly, we require the sources to have half-light radius (as measured via \emph{shape\_r} parameter in the Tractor catalog) $r_{\rm eff} > 2.5^{\prime\prime}$ to focus on the extended galaxies, and simultaneously  require $r_{\rm eff} < 20^{\prime\prime}$ to reject spurious sources or imaging artifacts, following \cite{2018ApJ...857..104G} and \cite{2021ApJS..252...18T} where the detections larger than this scale in HSC-SSP or DES images were inspected to be rare and generally spurious. By this criterion, 2,999,940 sources (out of 187,492,198) are retained.\par

Then, to avoid the sources seriously polluted by the nearby contaminants which would cause unreliable model fitting results, we require our sources to satisfy the following criteria according to \cite{2020RNAAS...4..187R}: 
\begin{equation}\label{eq1}
\begin{array}{ll}         \emph{fracmasked$\rm_{X}$} <0.5 \\
                          \emph{fracflux$\rm_{X}$} <5 \\
                          \emph{fracin$\rm_{X}$} >0.3 .
\end{array}
\end{equation}

where X represents the $g$, $r$, or $z$ bands. The \emph{fracmasked$\rm_{X}$}, \emph{fracflux$\rm_{X}$} and \emph{fracin$\rm_{X}$} are parameters in the Tractor catalogue which could probe the quality of the model fitting for the sources. Specifically, the \emph{fracmasked$\rm_{X}$}, the profile-weighted fraction of pixels masked from other observations of the target object, is used to remove sources with a high fraction of masked pixels. 
The \emph{fracflux$\rm_{X}$}, the profile-weighted fraction of the flux from other sources divided by the target object flux, is used to reject objects with heavily contaminated flux.
The \emph{fracin$\rm_{X}$}, the fraction of the flux from the target source within the blob, a group of pixels, is used to select sources with a large fraction to ensure well-constrained model fits.
By this criterion, 1,622,986 sources, approximately 0.446\% of the total sample, are retained.

Here before the next criterion, we correct the flux for the Galactic extinction and convert it to the magnitude with the prescription (Equation~\eqref{eq2}) given by the Legacy Surveys. 

\begin{equation}\label{eq2}
\begin{array}{ll}        m_{\rm {X}}{\rm = 22.5 - 2.5 log_{10}}(F_{\rm X})\\
                         F_{\rm corr,X} = F_{\rm X}/MW_{\rm X}\\
                         m_{\rm corr,X}{\rm = 22.5 - 2.5 log_{10}}(F_{\rm corr,X}) .
\end{array}
\end{equation}
where X represents the $g$, $r$, or $z$ bands. $F_{\rm X}$ is the model flux in the X band, measured as \emph{flux$_{\rm X}$} in unit of nanomaggy in the Tractor catalogue, $ MW_{\rm X}$ the Galactic transmission of the object position in the X filter, measured as \emph{mw$\_$transmission$\_$\rm X} in linear units from 0 to 1 in the catalogue, where 1 represents a fully transparent region of the Milky Way and 0 a fully opaque region. $F_{\rm corr,X}$, the Galactic-extinction corrected flux, is further converted to the magnitude, $m_{\rm corr,X}$, based on which the colors of $g$ - $r$ and $g$ - $z$ are obtained.

After that, we request the colors to be within the color box defined by 
\begin{equation}\label{eq3}
\begin{array}{ll}        -0.4 < g-z <2.3 \\
                    (g-r)< 0.6 \times (g-z) +0.6 \\
                    (g-r)> 0.6 \times (g-z) -0.1 .
\end{array}
\end{equation}

\begin{figure}[h]
   \centering
   \includegraphics[width=9cm, angle=0]{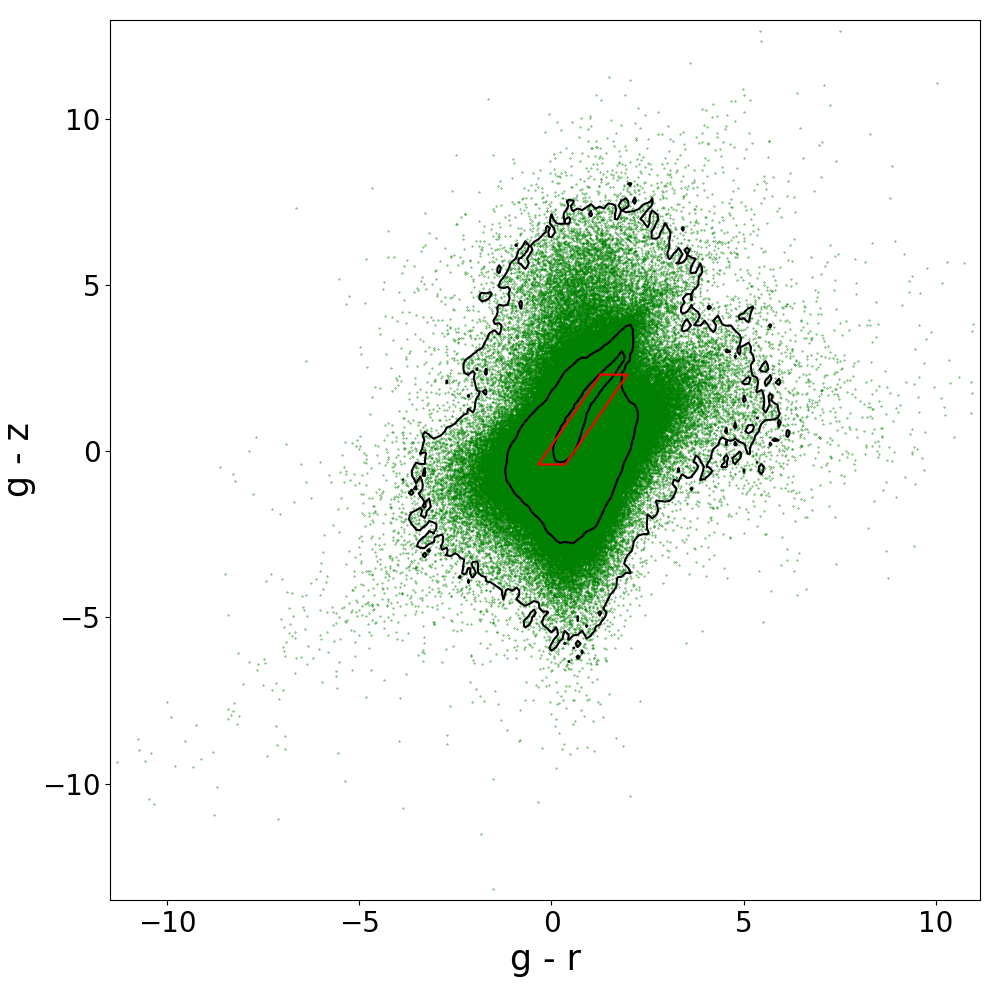}
   
   \caption{The $g$ - $r$ versus $g$ - $z$ diagram of the total sample of BASS+MzLS. The three black contours from the inside out are enclosing 68.2$\%$, 95.6$\%$, and 99.8$\%$ of the total sample. The red box is our color box expressed in Equation ~\eqref{eq3}, enclosing the most densely populated region of the galaxies while rejecting the high redshift galaxies and spurious objects.}
   \label{Fig1}
   \end{figure}

  This color box was empirically determined based on the distribution of the total sample of the BASS+MzLS in the $g$ - $r$ versus $g$ - $z$ diagram, as shown in Figure \ref{Fig1} where the three black solid contours from the inside out, respectively, enclose 68.2$\%$, 95.6$\%$, 99.8$\%$ of the total sample. For determining the color requirements (the red box in Figure \ref{Fig1}), our principles are including the majority of the galaxies within the central contour where 68.2$\%$ of the total sample gathers while excluding the high redshift galaxies and spurious objects.
  By satisfying the color requirements (equation~\eqref{eq3}), 994,459 sources ($\sim0.273\%$ of the total sample) are retained.\par

Subsequently, we require the ellipticity (1 - b/a) to be less than 0.7 (an axis ratio, b/a, greater than 0.3) to avoid edge-on galaxies, some spurious objects with high ellipticity (e.g., diffraction spikes), or the most obvious lensed galaxies. By this criterion, 772,745 sources (0.212\% of the total sample) are retained.
 
 Finally, we calculate the mean surface brightness within the half-light radius, $ \bar{\mu}_{\rm eff,X} $, by using Equation~\eqref{eq4}:
 \begin{equation}\label{eq4}
\begin{array}{ll}        
                         \bar{\mu}_{\rm eff,X} = 22.5 - 2.5{\rm log}_{10}(\frac{F_{\rm corr,X}}{2\pi r_{\rm eff}^{2}}) .
                         
\end{array}
\end{equation}
where $ r_{\rm eff} $ is the half-light radius, measured as \emph{shape\_r} in the Tractor catalogue. We require the $ \bar{\mu}_{\rm eff,X} $ to be within $24.2 < \bar{\mu}_{\rm eff,g} < 28.8$ mag arcsec$^{\rm -2}$ and obtain 344,370 objects (0.095\% of the total sample) as the initial sample of the LSBG candidates.  

For a clear picture of the process of our selection for the initial LSBG candidates so far, the selection criteria above are listed in Table \ref{selectstep}. Up to now, the selection of the initial LSBG candidates were solely via the direct use of the Tractor catalogue, so we furthermore inspected the images of a few thousand initial LSBG candidates and found a large number of the candidates were apparently false LSBGs that are instead the sources of contaminations. So it is necessary to reject those false LSBG candidates from the numerous initial candidates via the machine learning techniques.

\begin{table}
\bc
\begin{minipage}[]{100mm}
\caption[]{LSBG Selection Parameters\label{selectstep}}\end{minipage}
\setlength{\tabcolsep}{10pt}
\small
 \begin{tabular}{cccc}
  \hline\noalign{\smallskip}
Criterion& Range& LSBG Candidates& Percent\\
  \hline\noalign{\smallskip}
no cut & NA & 364,277,779 & 100.000\% \\
\emph{type} & $!=$ PSF, DEV, DUP & 187,492,198 & 51.470\% \\
$r_{\rm eff}$ & 2.5$^{\prime\prime}$ - 20$^{\prime\prime}$ & 2,999,940 & 0.824\% \\
\emph{fracmasked}, \emph{fracflux}, \emph{fracin} & Equation~\eqref{eq1} & 1,622,986 & 0.446\%\\
color & Equation~\eqref{eq3} & 994,459 & 0.273\%\\
ellipticity & $<$ 0.7 & 772,745 & 0.212\% \\
$\bar{\mu}_{\rm eff,g}(\rm mag$ $\rm arcsec^{\rm -2})$ & 24.2 - 28.8 & 344,370 & 0.095\%\\
Machine Learning & -  & 57,934 & 0.016\% \\
Visual Inspection & -  & 31,825 & 0.009\% \\
  \noalign{\smallskip}\hline
\end{tabular}
\ec

\end{table}

\subsection{Machine Learning Classification}
\begin{figure}	
	\centering
	\begin{subfigure}[t]{0.14\linewidth}
		\centering
		\caption{}\label{Fig2a}
		\includegraphics[scale=0.267]{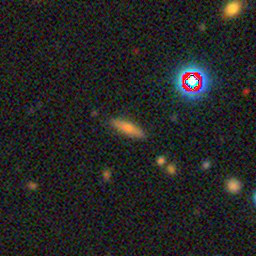}				
	\end{subfigure}
	\quad
	\begin{subfigure}[t]{0.14\linewidth}
		\centering
		\caption{}\label{Fig2b}
		\includegraphics[scale=0.267]{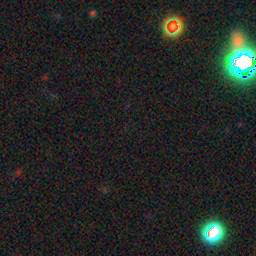}			
	\end{subfigure}
	\quad
	\begin{subfigure}[t]{0.14\linewidth}
		\centering
		\caption{}\label{Fig2c}
		\includegraphics[scale=0.267]{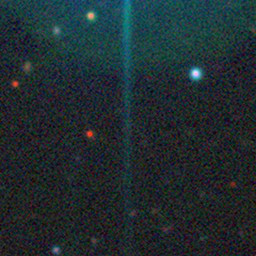}				
	\end{subfigure}
	\quad
	\begin{subfigure}[t]{0.14\linewidth}
		\centering
		\caption{}\label{Fig2d}
		\includegraphics[scale=0.267]{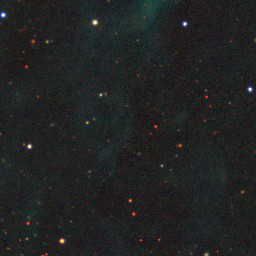}			
	\end{subfigure}
	\quad
	\begin{subfigure}[t]{0.14\linewidth}
		\centering
		\caption{}\label{Fig2e}
		\includegraphics[scale=0.267]{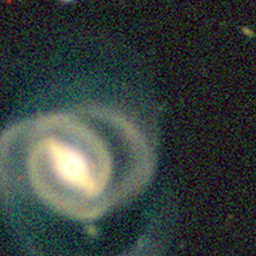}			
	\end{subfigure}
	\quad
	\begin{subfigure}[t]{0.14\linewidth}
		\centering
		\caption{}\label{Fig2f}
		\includegraphics[scale=0.267]{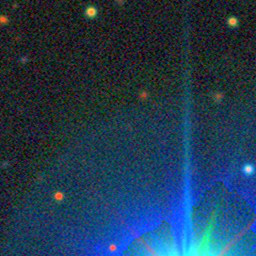}			
	\end{subfigure}

\caption{The composite images of the $g$, $r$, and $z$ bands from the Legacy Surveys DR9 for the common sources of contamination in the initial LSBG candidates. The size is $1.1^{\prime}\times1.1^{\prime}$ for all of the panels except for the panel \subref{Fig2d} which is of $3.84^{\prime}\times3.84^{\prime}$. The false LSBG candidates are at the center of each panel.}
\label{Fig2}
\end{figure}

From our visual inspection, the most common sources of contaminations for the false LSBGs were:
\begin{enumerate}
    \item Red objects with high ellipticity close to the criterion of 0.7 (e.g., Figure~2\subref{Fig2a}).
    \item Detections that are almost invisible in the images (e.g., Figure~2\subref{Fig2b}).
    \item Diffuse light from the nearby bright stars (e.g., Figure~2\subref{Fig2c} and ~2\subref{Fig2f}).
    \item Faint, diffuse regions of objects in a larger scale, such as Galactic cirrus (e.g., Figure~2\subref{Fig2d}).
    \item Diffuse light from the arms of large spiral galaxies (e.g., Figure~2\subref{Fig2e}).
\end{enumerate}

 Aiming to reject the false LSBGs from the initial sample of the LSBG candidates and simultaneously maintain a completeness of the true LSBGs as high as possible, we employed a supervised machine learning classification algorithms. 
 
\subsubsection{Training and test sets} \label{Training Set and Test Set}
\begin{figure}[h]
   \centering
   \includegraphics[width=14cm, angle=0]{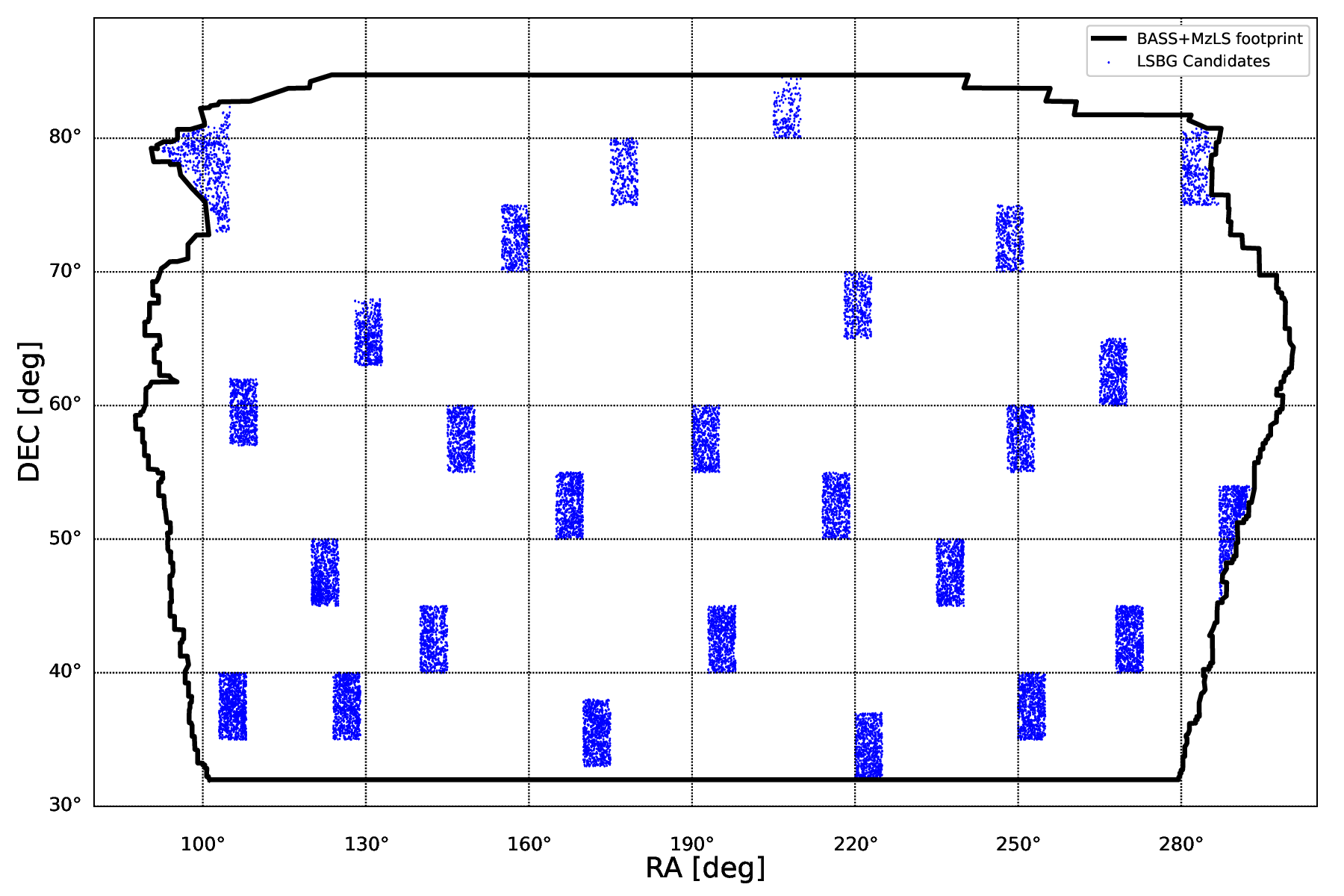}
   \caption{The 26 areas (blue) selected to generate a labeled set for training and test by the machine learning. The black solid outline encloses the entire sky area of the BASS+MzLS.}
   \label{Fig3}
   \end{figure}

In order to prepare for a labeled sample with objects labeled as either true or false LSBGs for the training and the test in machine learning, we decided to visually inspect the images of all of the 22,710 initial LSBG candidates within the 26 sky areas (blue areas in Figure \ref{Fig3}) that were selected by us to distribute uniformly in the spatial area of the BASS+MzLS. To alleviate the subjective biases, we had three individuals to perform the visual inspections independently to identify each candidate to be a true or a false LSBG. Then, the results from the three were combined as the final results. Ultimately, we labeled the 2,561 candidates identified as the true LSBGs by more than two individuals as LSBGs and labeled the rest 20,149 as non-LSBGs. Then, 70$\%$ of the labeled sample of 22710 labeled objects was adopted as a training set while the rest 30$\%$ of the labeled sample was utilized as a test set. We used the training set to train a model and  evaluated the quality of the trained model using the test set. \par

\subsubsection{Model, features and Classification}\label{subsec:model-features}
Before training the model, it’s key to pick up a machine learning algorithm. We tested and evaluated the widely used algorithms of Random Forest (via the Python library {\sc scikit-learn}; \cite{scikit-learn}), XGBoost (\cite{2016arXiv160302754C} ; via the Python library {\sc xgboost}), 
Naive Bayes, AdaBoost, K Nearest Neighbor, Decision Tree, Random Forest, Support Vector Machines, and SVM with radial basis function (RBF) kernel (via an automated toolkit {\sc auto-sklearn} that integrates diverse machine learning algorithms; \cite{feurer-neurips15a}).
Among these models, we voted for the XGBoost which stood out with the highest accuracy on the test set to be our machine learning model in this study. \par

Asides from the model, we need to opt for the useful features for learning. We performed tests and assessments for the quality of different feature combinations for learning by using the control variable method. If the accuracy of the model takes the first priority, we believe that it is best to use all of the following 24 features in learning, which are listed in an order of importance.\par

\begin{enumerate}
    \item The ellipticity of objects, 1 - b/a.
    \item The half-light radius, \emph{shape\_r}.
    \item The colors of \emph{$g$ - $r$}, \emph{$g$ - $z$}, and \emph{$r$ - $z$} derived from the Galactic extinction corrected magnitudes.
    \item The Galactic extinction corrected magnitudes in the $g$, $r$, and $z$ bands, \emph{mag\_corr}.
    \item The profile-weighted fraction of the flux from other sources divided by the total flux in the $g$, $r$, and $z$ bands, \emph{fracflux}.
    \item The fraction of a source's flux within the detection in the $g$, $r$, and $z$ bands, \emph{fracin}.
    \item The profile-weighted fraction of pixels masked from all observations of this object in the $g$, $r$, and $z$ bands, \emph{fracmasked}.
    \item The mean effective surface brightness in the $g$, $r$, and $z$ bands, \emph{mu\_mean}.
    \item The power-law index for the S$\rm \Acute{e}$rsic profile model, measured as \emph{sersic} in the Tractor catalogue.
    \item The central surface brightness in the $g$, $r$, and $z$ bands, \emph{mu\_0}, which is converted from \emph{mu\_mean} by the transforming prescription provided in \cite{2005PASA...22..118G}.
\end{enumerate}

As for the training, our principle was to obtain a model with the maximum value for the Recall parameter to make sure that the true LSBGs in the training sample could be retained in positive predictions as completely as possible while maintaining the Precision parameter (the proportion of true LSBGs in the predicted LSBGs) as high as possible. To evaluate the model at a balance between the Recall and Precision, the Fbeta-measure criterion was introduced as an evaluation metric, which represents the weighted harmonic mean of both the Precision and the Recall. In our principle, the Recall parameter should have a greater weight than the Precision, so we use beta = 2, a commonly used value, as the standard for the Fbeta-measure in model evaluation. With these guidelines, we trained the XGBoost model by using grid search and {\sc optuna}, a hyperparameter optimization framework \citep{optuna_2019}, to optimize the hyperparameters of XGBoost model. After thousands of optimizations, we finally derived the trained XGBoost model with the optimized hyperparameters, such as 
\emph{max\_depth} = 6, \emph{n\_estimators} = 337, \emph{learning\_rate} $\approx$ 0.09, \emph{subsample} $\approx$ 0.393, \emph{scale\_pos\_weight} = 8 and so on.\par 

\begin{figure}[h]
   \centering
   \includegraphics[width=10cm, angle=0]{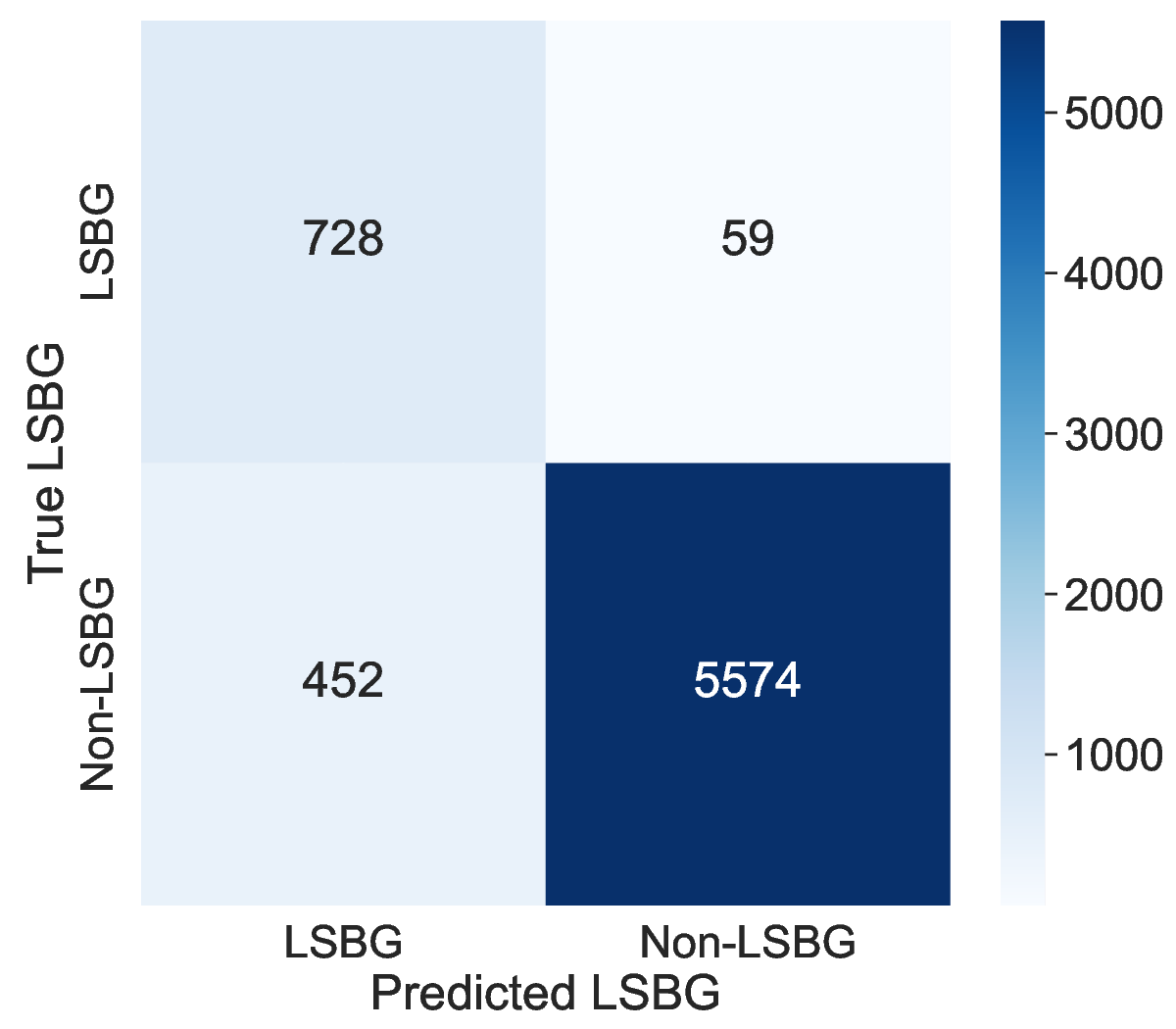}
   \caption{The confusion matrix of our XGBoost classifier evaluated on the test set. The quoted numbers correspond to the number of the test objects based on their true and predicted labels. The Recall is $\sim 92.5\%$.}
   \label{Fig4}
   \end{figure}

Subsequently, this XGBoost model was applied to the test set, and the results from the test set were displayed in the confusion matrix (Figure \ref{Fig4}). Obviously, the Recall value, defined as the ratio of the true LSBGs classified as LSBGs ($\rm Recall=TP/(FN+TP)$) by the model, is $ \sim 92.5\%$. For the minor fraction of the true LSBGs that were classified as non-LSBGs and the false LSBGs that were classified as LSBGs by the model, we visually inspected their images and found that they are too dark to result in a reliable classification. In addition, the Precision value, defined as the fraction of predicted LSBGs classified as ture LSBGs ($\rm Precision=TP/(FP+TP)$), is $\sim 61.7\%$, meaning that approximately 40\% of the objects in the LSBG candidates we obtained after machine learning are non-LSBGs. We validated this probability in Section \ref{Visual Inspection}.\par

With the help of the machine learning, the number of the initial LSBG candidates was decreased from 344,370 to 57,934. However, according to the Precision of the model, the 57,934 LSBG candidates are expected to still contain $\sim 40\%$ non-LSBGs, so we would perform the visual inspection of the images of the 57,934 candidates again to purify the sample in next section.

\subsection{Visual Inspection}\label{Visual Inspection}

In this section, we visually inspected the $grz$-composite images of the 57,934 LSBG candidates retained after the machine learning. 
From the inspection, we found that there are still false LSBGs in the sample whose visual appearances in the images were not like the true LSBGs at all, but the values of their main features listed in Section~\ref{subsec:model-features} given by the Tractor measurements followed the true LSBGs, making it challenging to classify them to be non-LSBGs by our model that were trained solely on learning the main features since we desired a fast learning and classification of the LSBGs in this work. However, in the future, we plan to train a better deep learning classification model using both the features and images of the final LSBG sample selected in this work. 

Specifically, these false LSBGs in the current sample still appeared to be like the contaminations shown in Figure \ref{Fig2}.
Therefore, we rejected them by visual inspection and ultimately resulted in a final sample of 31,825 LSBG candidates with a high purity of the true LSBGs with the half-light radius 2.5$^{\prime\prime}$ $< r_{\rm eff} <$ 20$^{\prime\prime}$ and the Galactic extinction-corrected mean effective surface brightness $24.2 < \bar{\mu}_{\rm eff,g} < 28.8$ mag arcsec$^{\rm -2}$. This final sample is so far the largest catalogue of LSBG candidates from the $\sim$ 5500 deg$^{2}$ sky area of BASS+MzLS, more than 1/3 of the entire sky area of the DR9 of the DESI Legacy Survey. 


\section{LSBG PROPERTIES}
\label{LSBG PROPERTIES}
We successfully established a sample of 31,825 LSBG candidates from the BASS+MzLS, spanning a wide range of properties, such as the color, morphology and environment, which would be studied in detail in this section.

\subsection{Color Distribution}\label{subsec:color}

   \begin{figure}[h]
   \centering
   \includegraphics[width=14cm, angle=0]{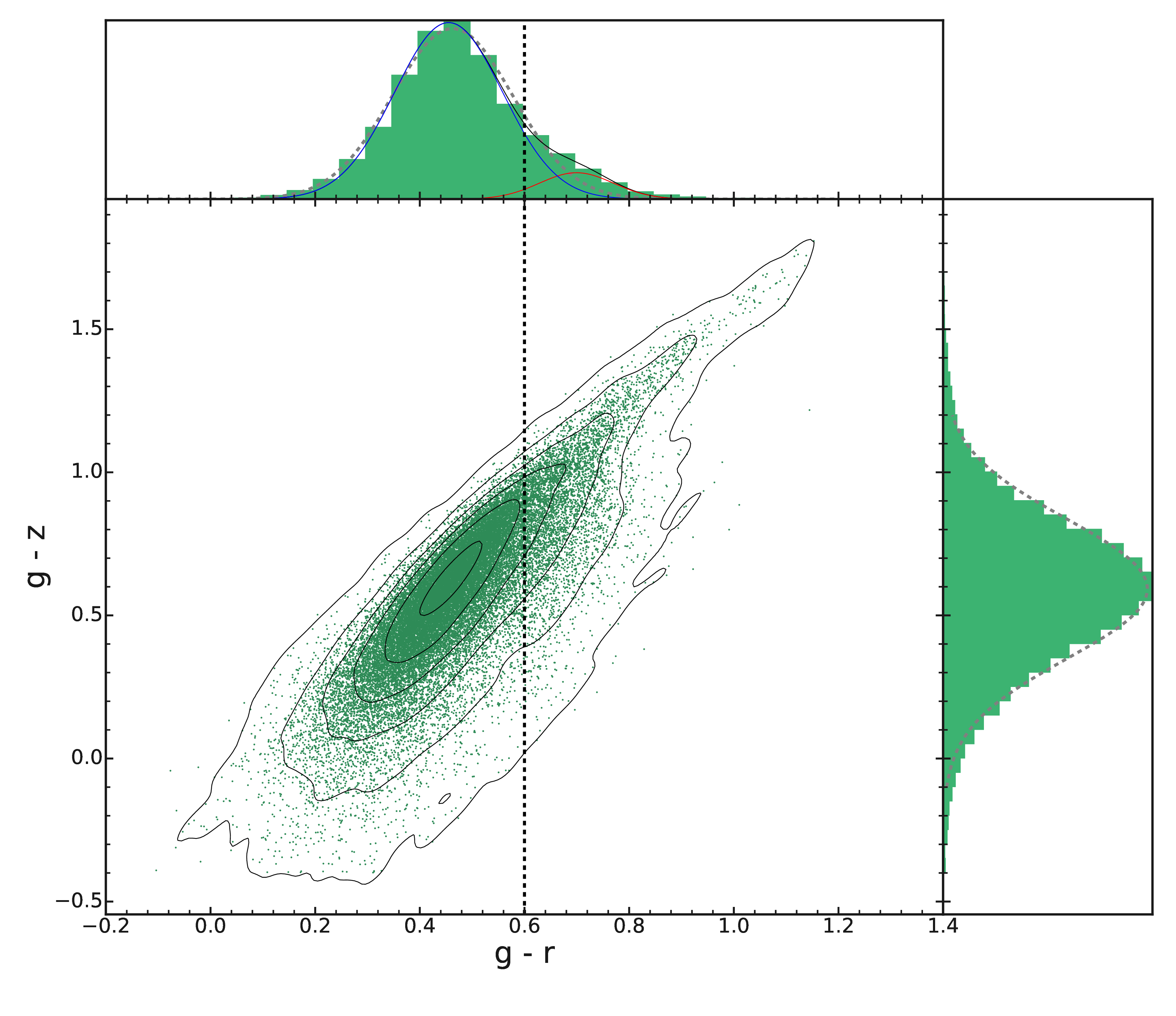}
   \caption{The diagram of $g$ - $r$ versus $g$ - $z$ for the final sample LSBG candidates (green) and the contours from the kernel density estimates (black isochrones). The top and the right panels show the histogram of $g$ - $r$ and $g$ - $z$, respectively. In the top panel, the $g$ - $r$ distribution is best fitted by the sum (black solid profile) of the two single Gaussian profiles (blue and red solid curves). By comparison, the grey dashed curve represents the fitted single Gaussian profile which is abandoned according to the evaluation by the AIC/BIC. The vertical black dashed line at $g$ - $r$ = 0.60 is the dividing line to distinguish red ($g$ - $r$ $>$ 0.60) from blue ($g$ - $r$ $<$ 0.60) LSBG candidates.}
   \label{Fig5}
   \end{figure}

We displayed the distribution of the final sample of the LSBG candidates in the color - color diagram of $g$ - $r$ versus $g$ - $z$ in Figure \ref{Fig5}. The sample galaxies (green dots) exhibited a bimodal distribution in the $g$ - $r$ color which naturally required a fitting by a combination of double Gaussian profiles rather than a single Gaussian profile according to our evaluation by the Akaike Information Criterion (AIC / AICc) and the Bayesian Information Criterion (BIC; see details in Section \ref{subsec:discussion-fitting}). 

In Figure \ref{Fig5} (the top panel), the best-fitting profile (black solid curve) evaluated by the AIC/BIC is the sum of a blue component represented by a single Gaussian profile with a peak value at a blue $g$ - $r$ color of 0.455 and $\sigma$ of 0.103 (blue solid curve) and a red component represented by a single Gaussian profile with a peak value at a red $g$ - $r$ color of 0.700 and $\sigma$ of 0.070 (red solid curve). Obviously, the blue component is dominated by the blue LSBG candidates of which 97.8$\%$ are bluer than $g$ - $r$ $<$ 0.66 while the red component is dominated by the red LSBG candidates of which 97.8 $\%$ are redder than $g$ - $r$ $>$ 0.56. This means that galaxies between $g$ - $r$ = 0. 56 and 0.66 are the mixture of LSBG candidates from the red end of the blue component ($g$ - $r$ $>$ 0.56) and those from the blue end of the red component ($g$ - $r$ $<$ 0.66). Since the median color of all of the galaxies between 0. 56 and 0.66 is 0.60 in $g$ - $r$, we adopt $g$ - $r$ = 0.60 as the color dividing line (vertical black dashed line) to separate the final sample of LSBG candidates into two subsamples of the blue ($g$ - $r$ $<$ 0.60; 26,672 galaxies) and the red ($g$ - $r$ $>$ 0.60; 5,153 galaxies). The median $g$ - $r$ colors of the blue and red subsamples are 0.44 and 0.67, respectively. 

\begin{figure}
    \centering
    \subcaptionbox{}{\includegraphics[width=.48\linewidth]{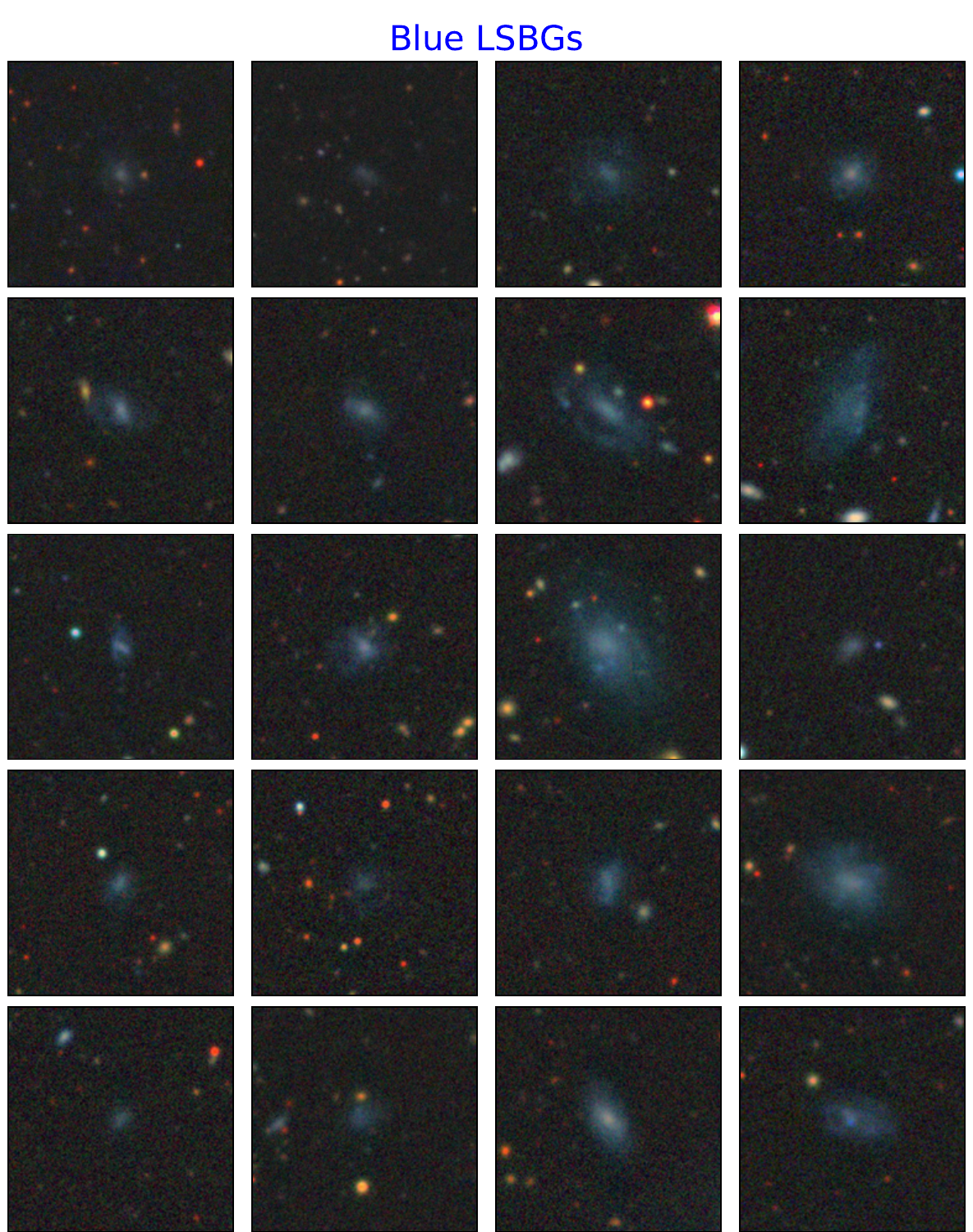}}\hspace{4pt}
    \subcaptionbox{}{\includegraphics[width=.48\linewidth]{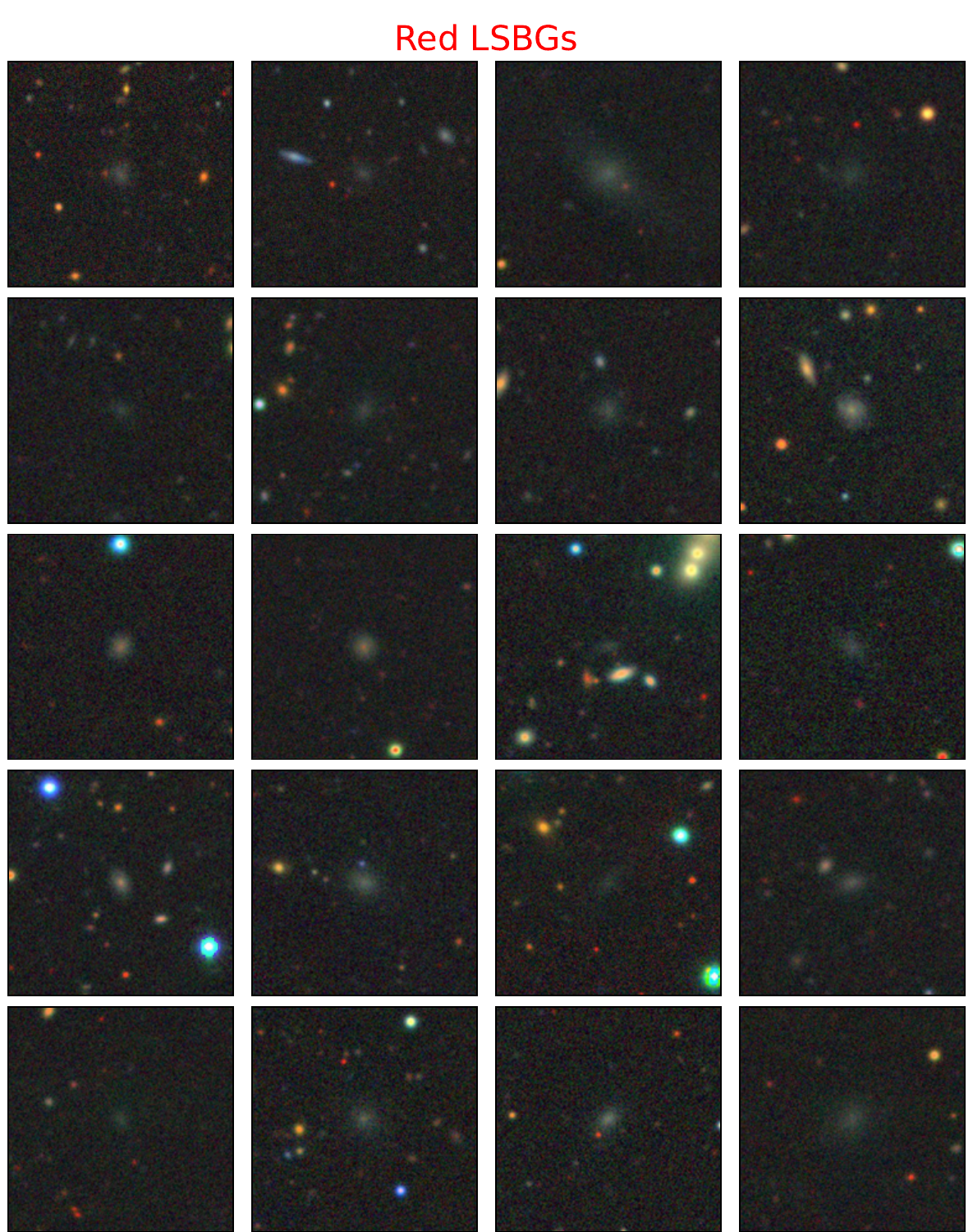}}
    \caption{The $grz$-composite images from the DESI Legacy Survey DR9 for several example LSBG candidates from the blue (left) and red (right) subsamples. The frame size for each LSBG candidate is $1.1^{\prime} \times 1.1^{\prime}$.}
    \label{Fig6}
\end{figure}

In Figure~\ref{Fig6}, we show randomly selected LSBG candidates from the the blue (the left) and the red (the right) subsamples for examples. Apparently, the blue LSBG candidates appear disk-like, spiral or irregular while the red ones tend to be spheroidal or elliptical. The former is quite distinguishing from the latter in morphology, implying that the colors of LSBGs correlate with their morphologies. Such a conclusion was also supported by several previous published studies, which would be discussed in Section \ref{subsec:discussion-comp}. \par

\subsection{Magnitude and Surface Brightness}

\begin{figure}
    \centering
    \subcaptionbox{}{\includegraphics[width=.31\linewidth]{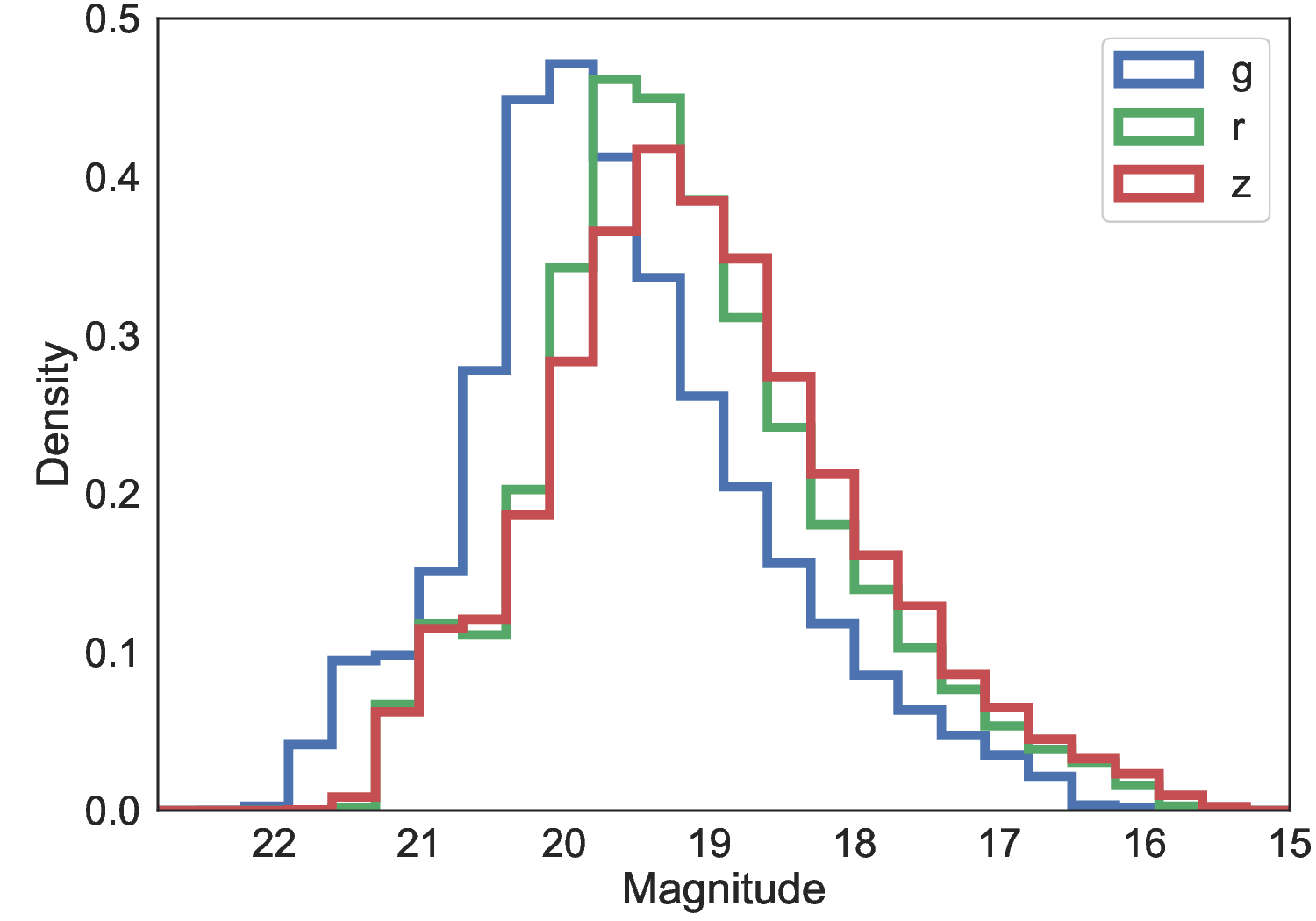}}\hspace{4pt}
    \subcaptionbox{}{\includegraphics[width=.31\linewidth]{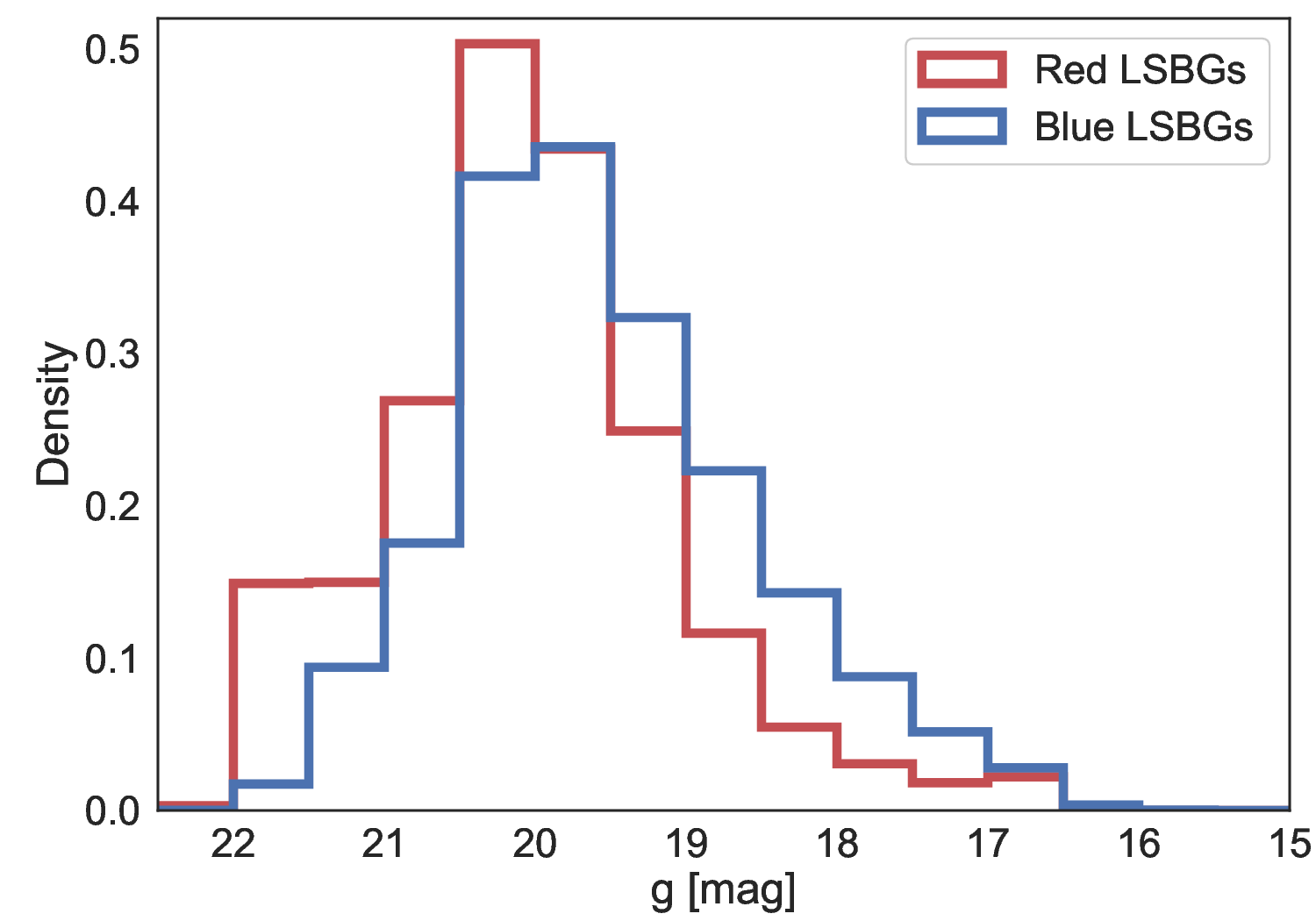}}\hspace{4pt}
    \subcaptionbox{}{\includegraphics[width=.31\linewidth]{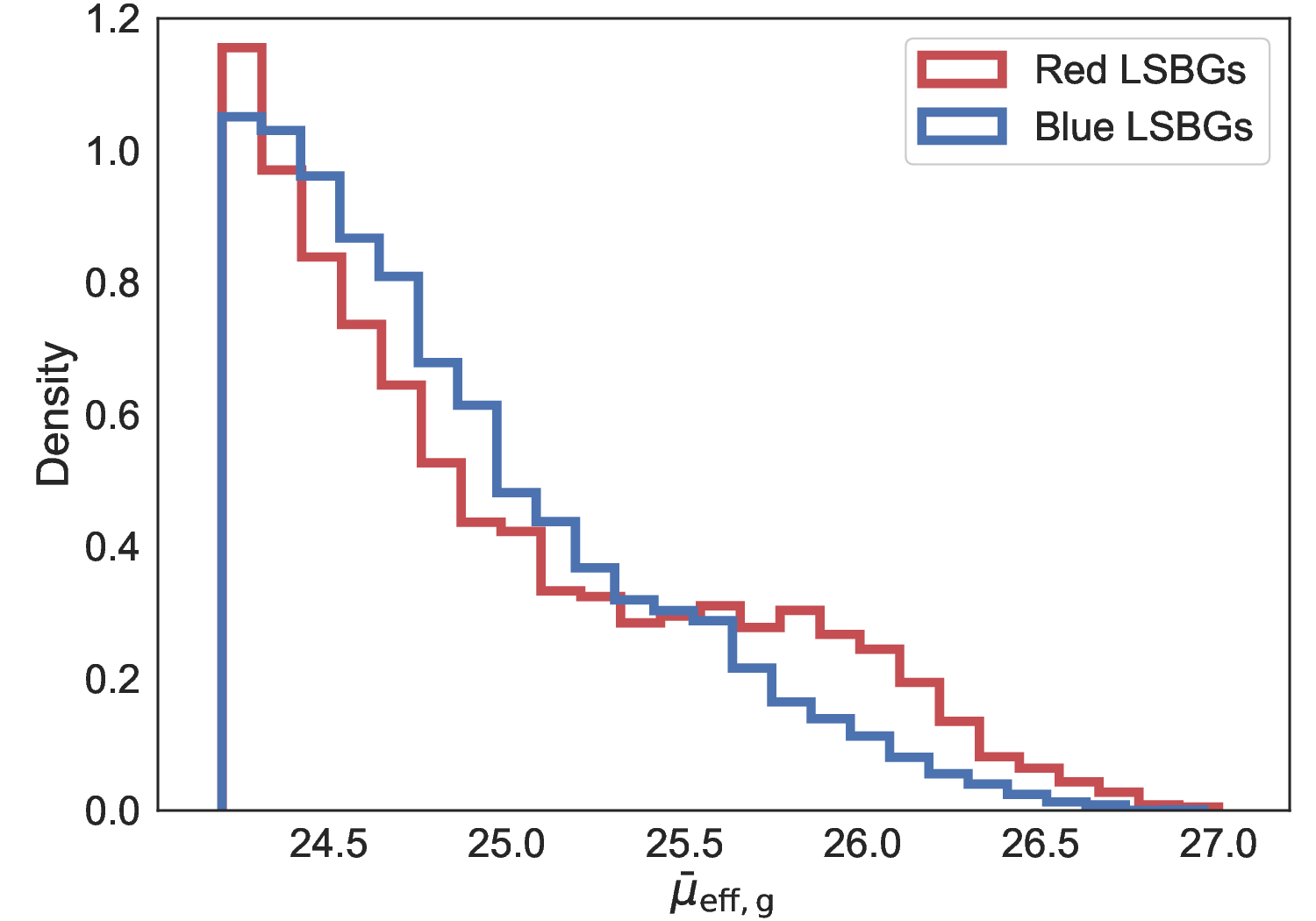}}
    \caption{The distribution of $g$-, $r$-, and $z$-band magnitudes of the final sample of LSBG candidates (a). The $g$-band magnitude (b) and mean surface brightness $\bar{\mu}_{\rm eff,g}$(c) are displayed for the blue and the red subsamples, respectively.}
    \label{Fig7}
\end{figure}

In Figure \ref{Fig7}(a) the distribution of the magnitudes in the $g$-, $r$-, and $z$-band are shown for the final sample of LSBG candidates entirely. In Figure \ref{Fig7}(b), the distribution of the $g$-band magnitude is compared between the blue and red subsamples, showing that the blue are slightly brighter than the red in the apparent magnitude in $g$-band. 
In Figure \ref{Fig7}(c), we show the distribution of the mean surface brightness $\bar{\mu}_{\rm eff,g}$ for the blue and red subsamples, respectively. We find that the red subsample shows a bump or an excess at the lower surface brightness tail (fainter than 25.5 $g$ mag arcsec$^{-2}$) while the blue subsample has slightly more LSBGs with higher surface brightness (brighter than 25.5 $g$ mag arcsec$^{-2}$), implying that the red LSBGs from our sample incline to have lower surface brightness while the blue ones tend to have higher surface brightness. This could be further supported by the statistics that the 16$^{th}$, 50$^{th}$, and 84$^{th}$ percentiles of $\bar{\mu}_{\rm eff,g}$ are 24.4, 24.7, 25.5 mag arcsec$^{-2}$ for the blue subsample and 24.4, 24.8, 25.8 mag arcsec$^{-2}$ for the red subsample.\par

\subsection{Ellipticity, Effective Radius and S$\rm \Acute{e}$rsic Index}
\begin{figure}
    \centering
    \subcaptionbox{}{\includegraphics[width=.47\linewidth]{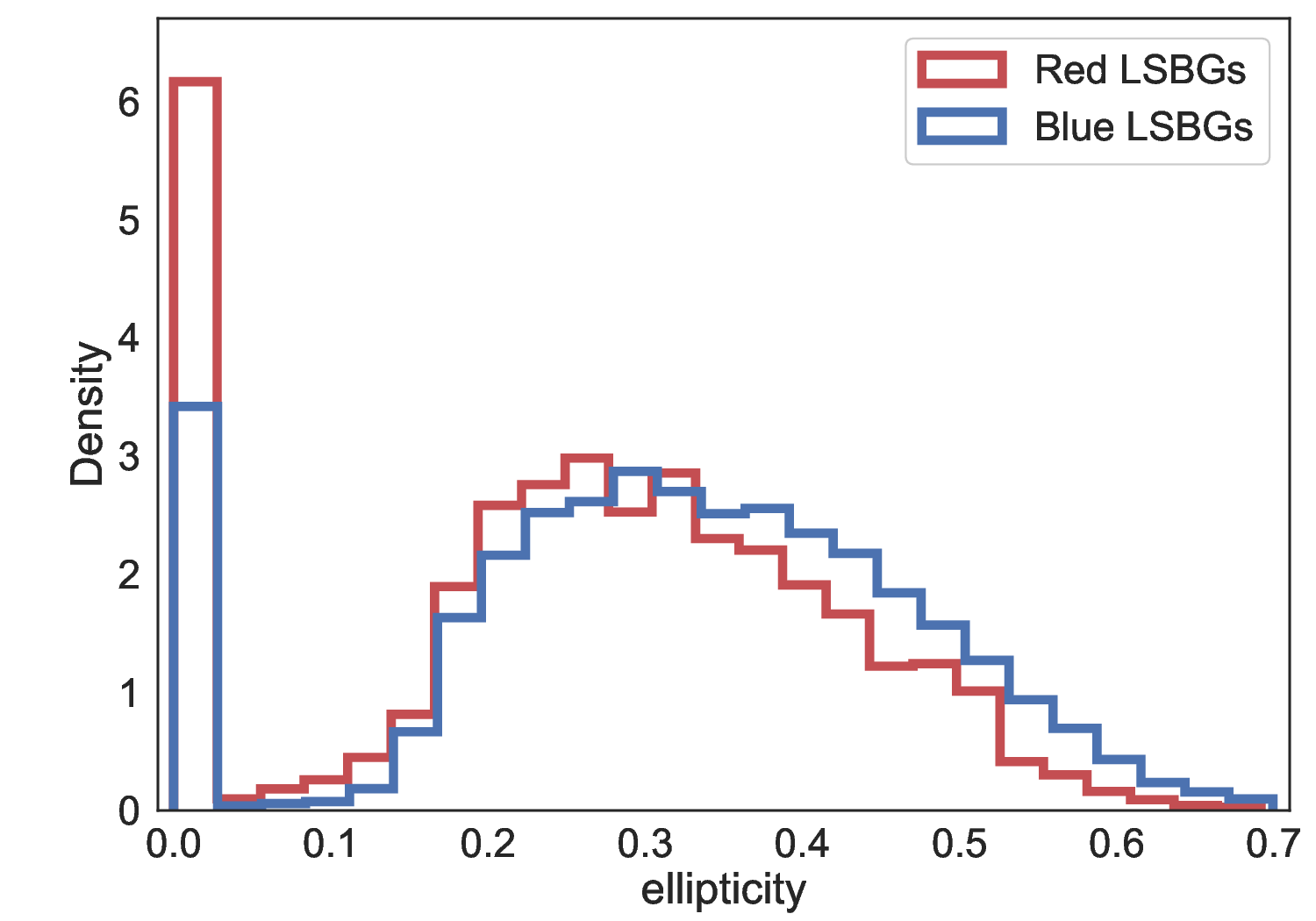}}\hspace{18pt}
    \subcaptionbox{}{\includegraphics[width=.47\linewidth]{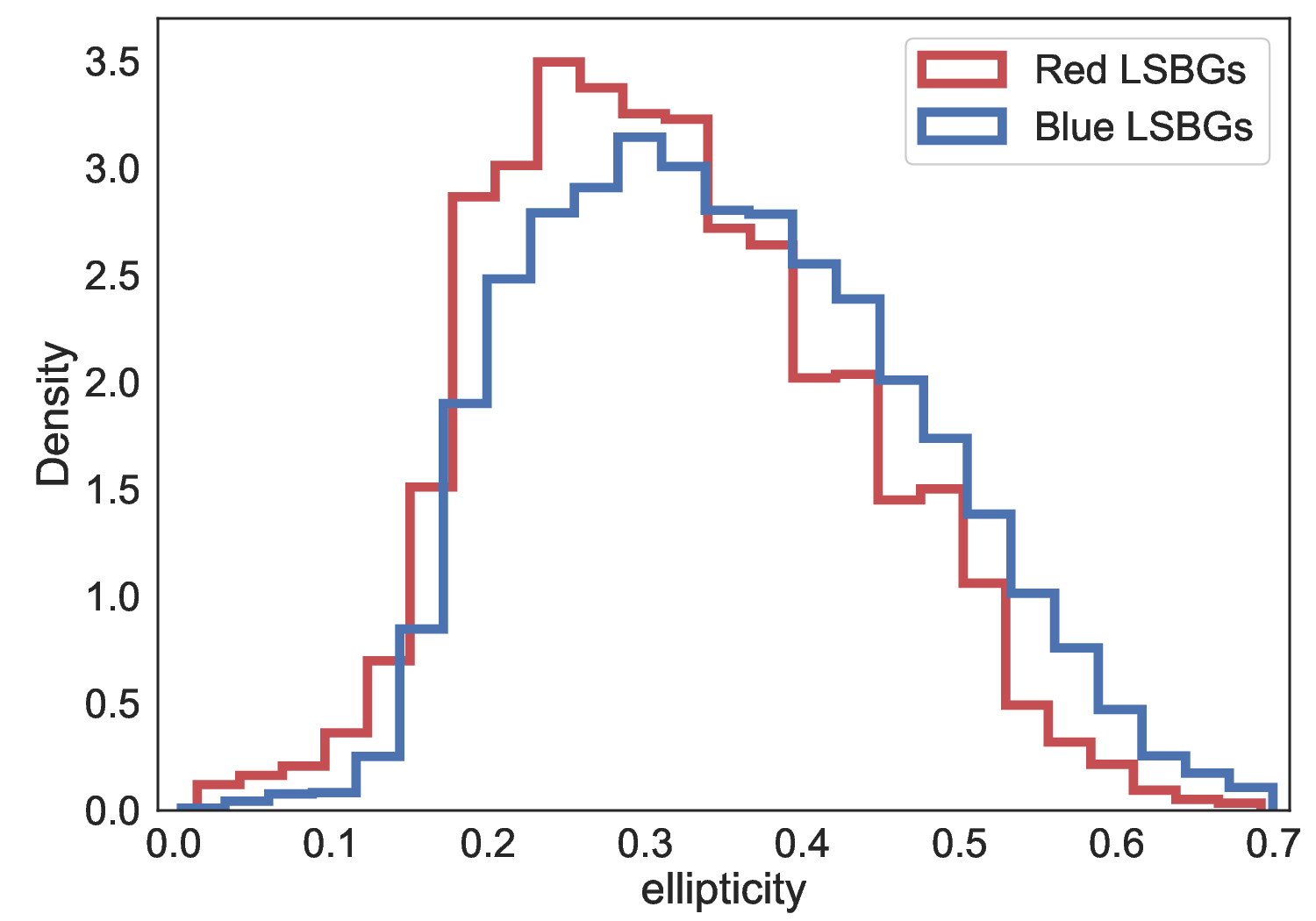}}
    \caption{The distribution of the ellipticity of the final sample of LSBGs are displayed for the blue and the red subsamples, respectively, in panel (a), where it shows the majority of the LSBGs have zero ellipticity from the Tractor catalogue. In panel (b), we exclude those galaxies with zero ellipticity only to give a clear picture of the distribution for the galaxies with non-zero ellipticity.}
    \label{Fig8}
\end{figure}

In Figure \ref{Fig8}(a), we present the distribution of ellipticity ($\epsilon$ = 1 - b/a) for the full final sample. It shows that both the blue and the red subsamples have considerable fractions of galaxies with the zero $\epsilon$ from the Tractor catalogue (10\% of the blue and 18\% of the red). The median $\epsilon$ are $\sim$ 0.31 for the full sample, $\sim$ 0.32 for the blue and $\sim$ 0.28 for the red, respectively. To give a clear picture of the $\epsilon$ distribution for those galaxies without the zero $\epsilon$, we plot a zoom-in picture for them in panel (b), where both subsamples show generally consistent distributions, with the median $\epsilon$ of 0.34 for the blue and 0.31 for the red, respectively. All of these $\epsilon$ distributions demonstrate that the LSBG candidates in our final sample are obviously round between $\epsilon$ = 0.1 and 0.7, which differs from the normal spiral galaxies showing a nearly flat ellipticity distribution between $\epsilon$ = 0.1 and 0.7 (Figure 4 in \cite{2013MNRAS.434.2153R}). 

\begin{figure}
    \centering
    \subcaptionbox{}{\includegraphics[width=.47\linewidth]{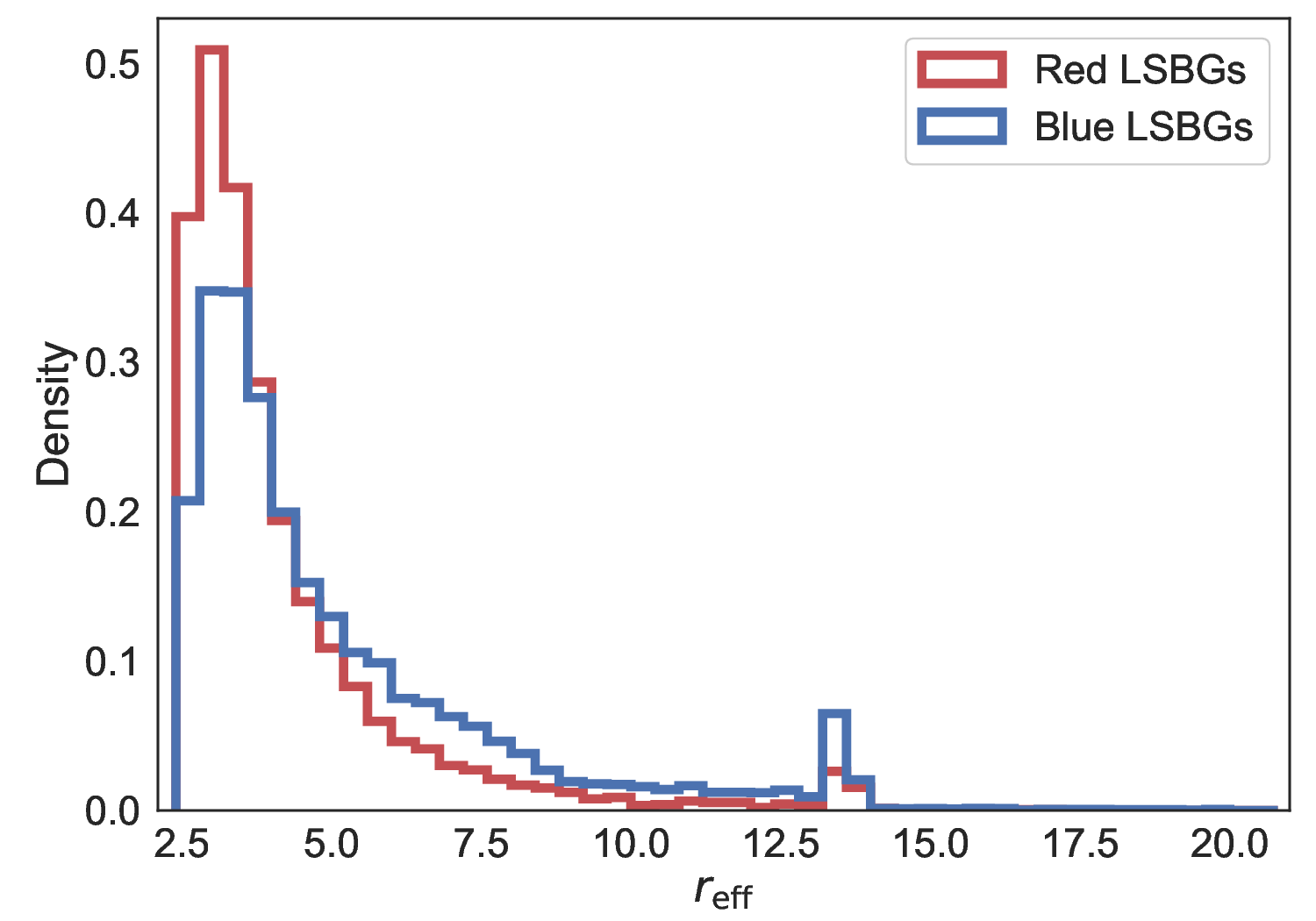}}\hspace{18pt}
    \subcaptionbox{}{\includegraphics[width=.47\linewidth]{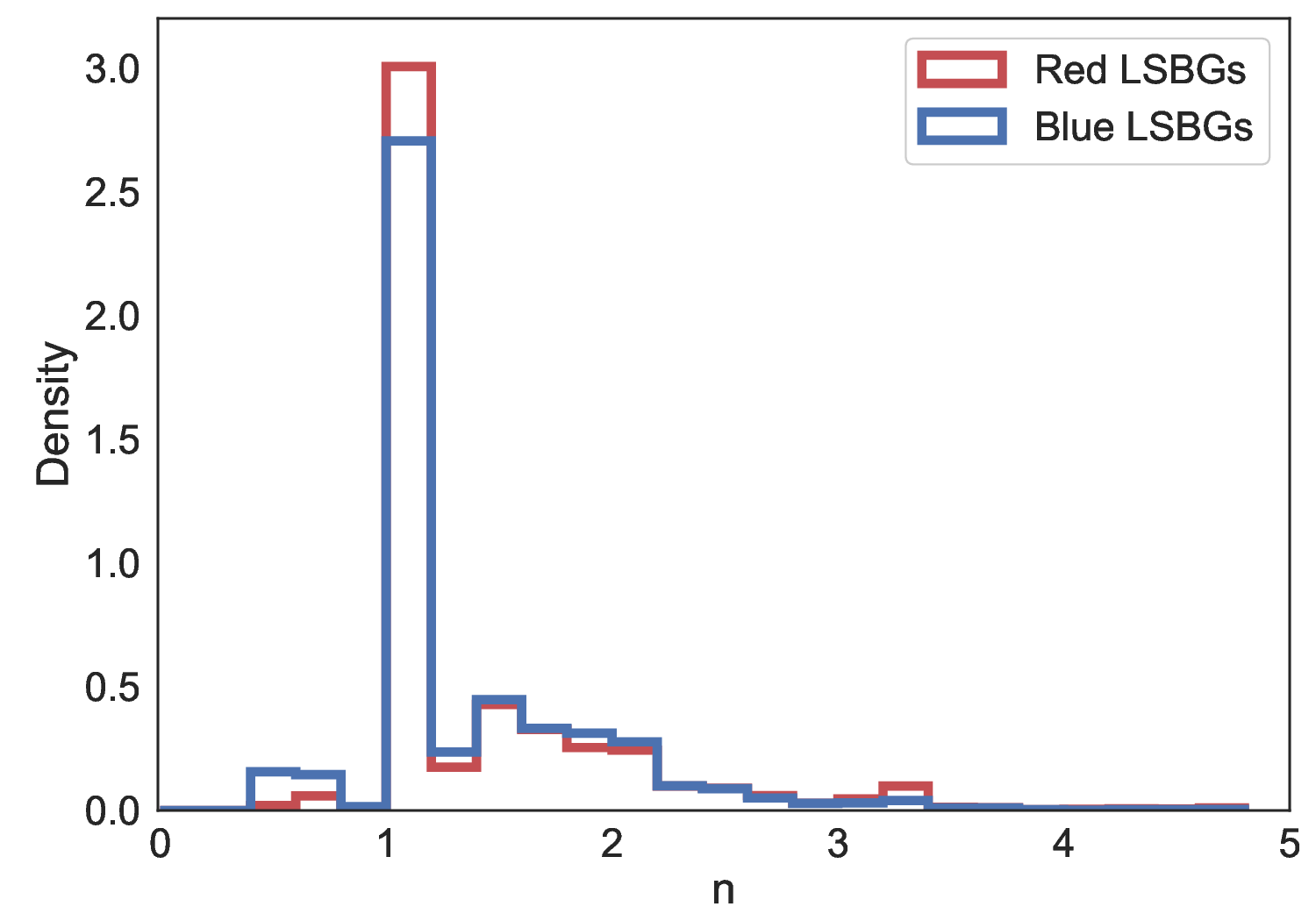}}
    \caption{The distribution of the half-light radius $r_{\rm eff}$ (left) and the S$\rm \Acute{e}$rsic index $n$ (right) for the blue and the red subsamples.}
    \label{Fig9}
\end{figure}

In Figure~\ref{Fig9}(a), both subsamples are dominated (more than 99\%) by galaxies with sizes ranging from 2.5$^{\prime\prime}$ to  14$^{\prime\prime}$ in $r_{\rm eff}$, with the medians are 3.5$^{\prime\prime}$ for the red sample, 4.1$^{\prime\prime}$ for the blue sample, and 4$^{\prime\prime}$ for the full sample. It is worth noting that the $r_{\rm eff}$ measurements from the Tractor catalogue for the minority of the large spiral galaxies are all given to be around $\sim$ 13.8$^{\prime\prime}$, causing a 
low peak occurs at $r_{\rm eff}$ of $\sim$ 13.8$^{\prime\prime}$ in the figure. This low peak due to the limitation of the Tractor model measurements has no physical implications, but the galaxies in this low peak all appear blue, large, diffuse, and extended disk LSBGs from our visual inspection. Thus, we still kept these galaxies in our final sample. In Figure \ref{Fig9}(b), we plot the S$\rm \Acute{e}$rsic index $n$ for the blue and the red subsamples, showing that 95\% of the blue subsample have $n<$ 2.5 while 93\% of the red subsample have $n<$ 2.5. The distribution of $n$ agrees with each other for both subsamples, with a median of $n$ = 1 for each, showing our final sample is dominated by the disk LSBGs. \par

\subsection{Spatial Distribution}
\begin{figure}[h]
       \centering
        \subcaptionbox{}{ \includegraphics[width=13cm, angle=0]{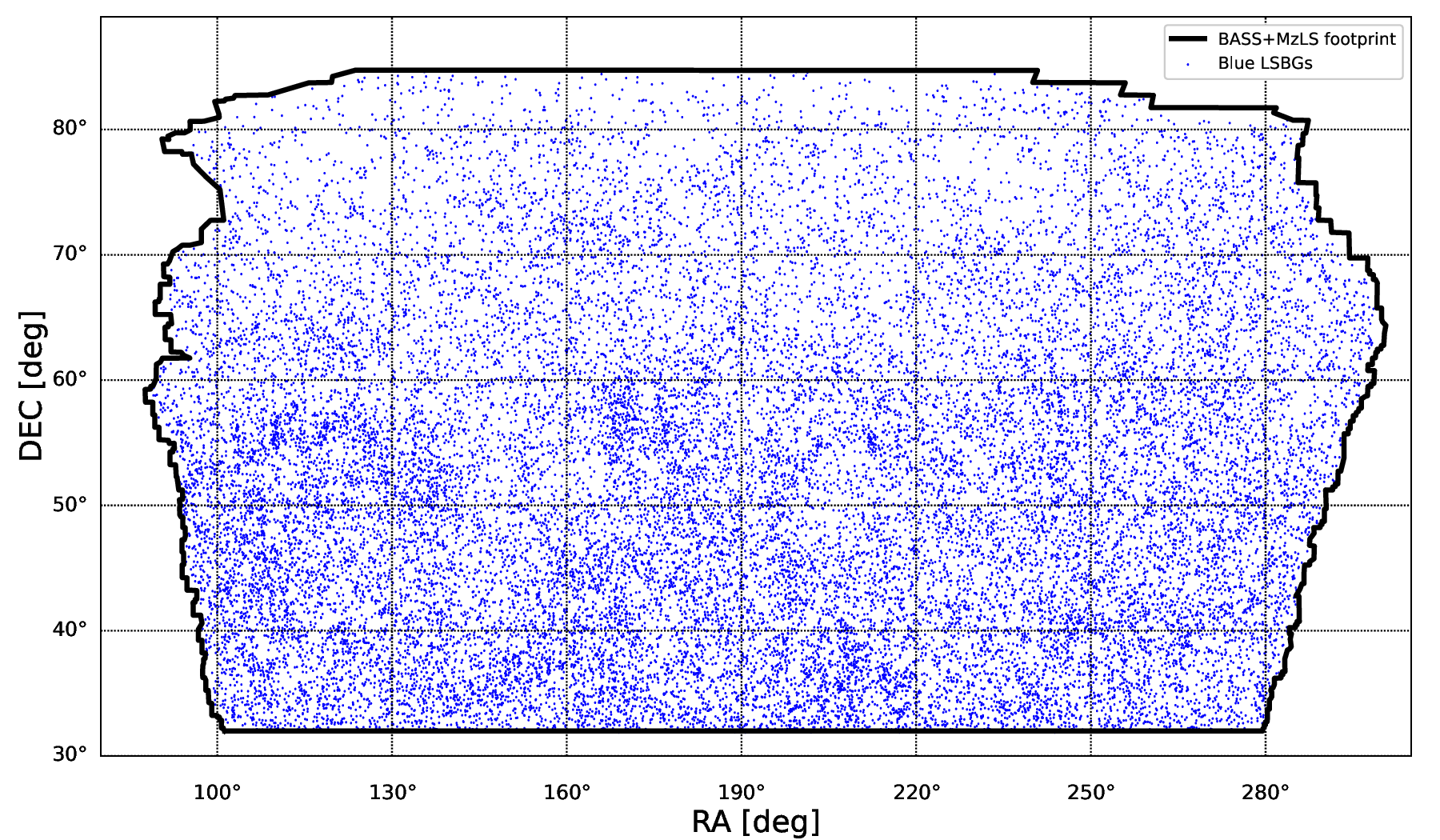}}
       \quad
        \subcaptionbox{}{ \includegraphics[width=13cm, angle=0]{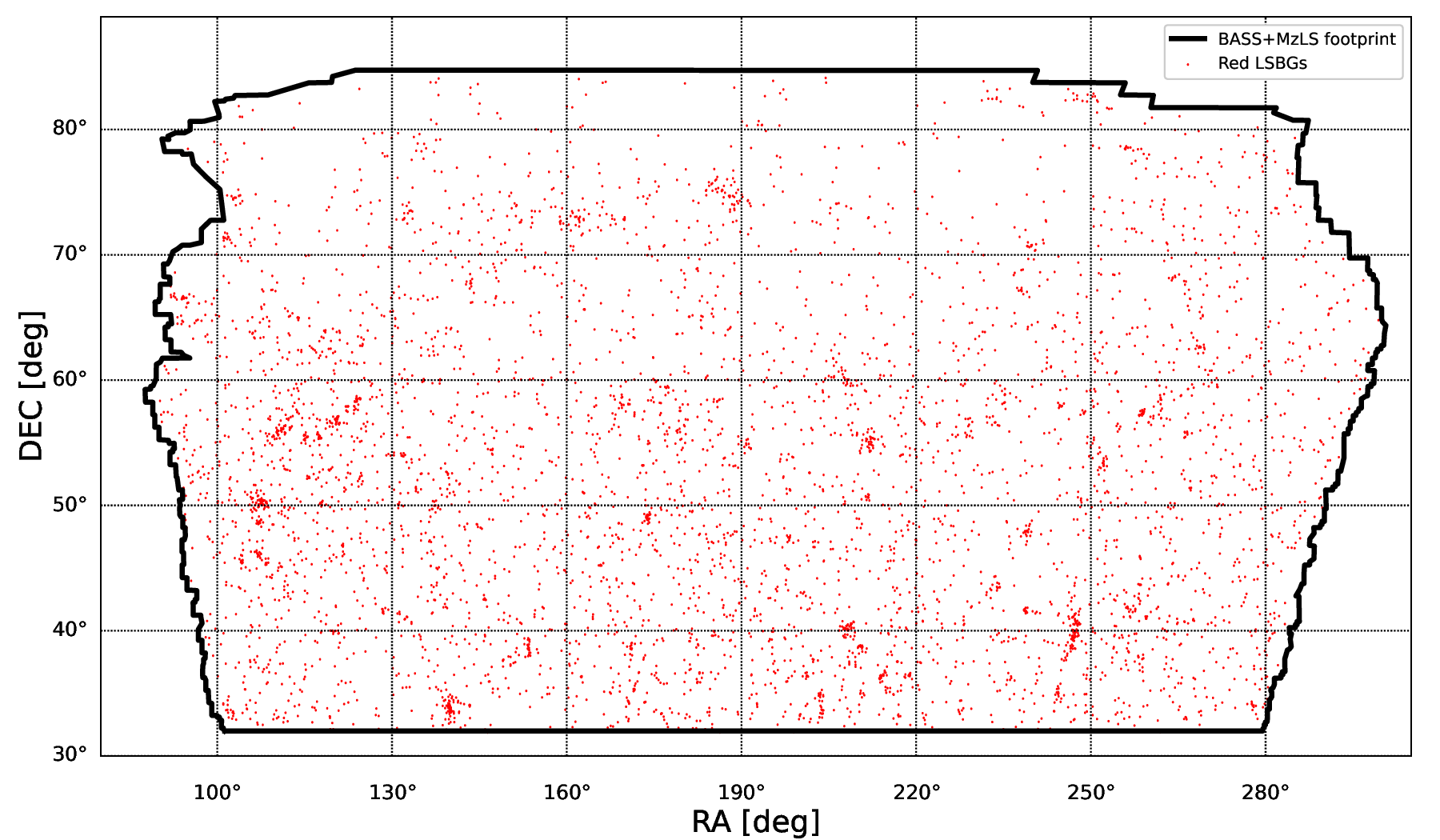}}
   \caption{The spatial distributions of the blue (top) and the red (bottom) subsamples of LSBGs within the footprint of BASS+MzLS survey (black solid).}
   \label{Fig10}
   \end{figure}

In Figure~\ref{Fig10} we show the spatial distribution of the blue (top) and the red (bottom) subsamples over the sky area within the the BASS+MzLS footprint (the black solid). We find an obvious discrepancy between the spatial distribution of the two subsamples. The blue LSBGs are more uniformly distributed while the red populations are clustered, showing that red LSBGs preferentially inhabit in denser environments than blue LSBGs. This is found by the studies of \cite{2018ApJ...857..104G} and \cite{2021ApJS..252...18T} as well, and we will discuss it in Section \ref{subsec:discussion-comp}.\par

\section{DISCUSSION}
\label{sect:discussion}

\subsection{Double or single Gaussian fitting?}\label{subsec:discussion-fitting}
In Section \ref{subsec:color} our LSBG sample was reported to have a bimodal color distribution that could be best fitted by a mixture of double Gaussian models rather than a single Gaussian model. Such a statement is supported by the evaluation of the performance of the single Gaussian model (SGM; grey dashed line in the top panel of Figure \ref{Fig5}) and the double Gaussian model (DGM; black line in the top panel of Figure \ref{Fig5}) fit according to the AIC/BIC in the equation below (equation~\eqref{eq5}) .

\begin{equation}\label{eq5}
\begin{array}{ll}        AIC=2k - 2\ln{\hat{L}} \\
                    AICc=AIC+(2k^2+2k)/(n-k-1)\\
                    BIC=\ln{(n)}k - 2\ln{\hat{L}} .
\end{array}
\end{equation}
where $k$ is the number of fitting parameters, $\hat{L}$ is the likelihood function, and n is the number of samples. When the sample is small in size, AIC should be corrected into AICc. According to \cite{doi:10.1080/01621459.1995.10476572}, the performance of the model improves as the AICc or BIC value decreases. 

We derive the BIC or AICc values from fitting the $g$ - $r$ color distribution of our sample with a single Gaussian model as BIC$_{\rm SGM}$ or AICc$_{\rm SGM}$. Similarly, we derive BIC$_{\rm DGM}$ or AICc$_{\rm DGM}$ for the fit with the double Gaussian model. Then, the BIC or AICc differences between the SGM and DGM are calculated as $\Delta BIC= BIC_{\rm SGM}-BIC_{\rm DGM}$ and $\Delta AICc= AICc_{\rm SGM}-AICc_{\rm DGM}$.
According to \cite{doi:10.1080/01621459.1995.10476572}, if $\Delta BIC$ or $\Delta AICc$ was larger than 10, the DGM would prevail. In our calculation, the $\Delta AICc$ and $\Delta BIC$ values are 235.1 and 227.1, respectively, which are far greater than 10, giving a strong evidence for us to believe that the $g$ - $r$ color distribution is much better fitted by a double Gaussian model than a single Gaussian model. This strongly convinces us of a bimodal $g$ - $r$ color distribution of the final sample of LSBGs. Additionally, such bimodal distributions of the colors of the LSBGs have also been reported for the previously defined sample of LSBGs from \cite{2018ApJ...857..104G} and \cite{2021ApJS..252...18T}, which would be discussed in detail in Section \ref{subsec:discussion-comp}.   \par

\subsection{Comparison with previous samples}\label{subsec:discussion-comp}

In this section, we compare our sample of the LSBG candidates with three other LSBG samples from \cite{2015AJ....149..199D}(D15), \cite{2018ApJ...857..104G}(G18) and \cite{2021ApJS..252...18T}(T21), respectively. 
The D15 provides a sample of 1,129 LSBGs selected from the 2800 deg$^{2}$ area of the $\alpha$.40 - SDSS DR7 survey with an imaging depth of $r \sim$ 22.2 mag for point sources of 95\% detection \citep{2000AJ....120.1579Y}. This sample is defined on the central surface brightness $\mu_{\rm 0,B} >$ 22.5 mag arcsec$^{-2}$, and they are nearby (z $<$ 0.06), blue, H{\sc{i}}-rich, and disk-dominated. The G18 presents a sample of 781 extended LSBGs from the first $\sim$ 200 deg$^{2}$ area of the imaging survey of the Wide layer of the Hyper Suprime-Cam Subaru Strategic Program (HSC-SSP) which has a depth of $g\sim$ 26.8, $r\sim$ 26.4, and $i \sim$ 26.4 mag for point sources at 5$\sigma$ \citep{2018PASJ...70S...8A}. This sample is defined on the mean surface brightness ($\bar{\mu}_{\rm eff,g} >$ 24.3 mag arcsec$^{-2}$) to allow nucleated galaxies into the sample and on galaxy size  ($r_{\rm eff} >$ 2.5$^{\prime\prime}$) as well to be restricted to low redshift. Using the similar selection criteria to the G18, the T21 produces a catalogue of 23,790 extended LSBGs from the $\sim$ 5000 deg$^{2}$ area of the first three years of imaging data from the Dark Energy Survey (DES Y3) with a depth of $g \sim$ 23.52, $r \sim$ 23.10, and $i \sim$ 22.51 mag for point sources at 10$\sigma$ which is corresponding to a surface brightness limit at 3$\sigma$ of $g \sim$ 28.26$^{+0.09}_{- 0.13}$, $r \sim$ 27.86$^{+0.10}_{- 0.15}$, and $i \sim$ 27.37$^{+0.10}_{- 0.13}$ mag arcsec$^{-2}$.\par

\begin{figure}
    \centering
    \subcaptionbox{}{\includegraphics[width=.31\linewidth]{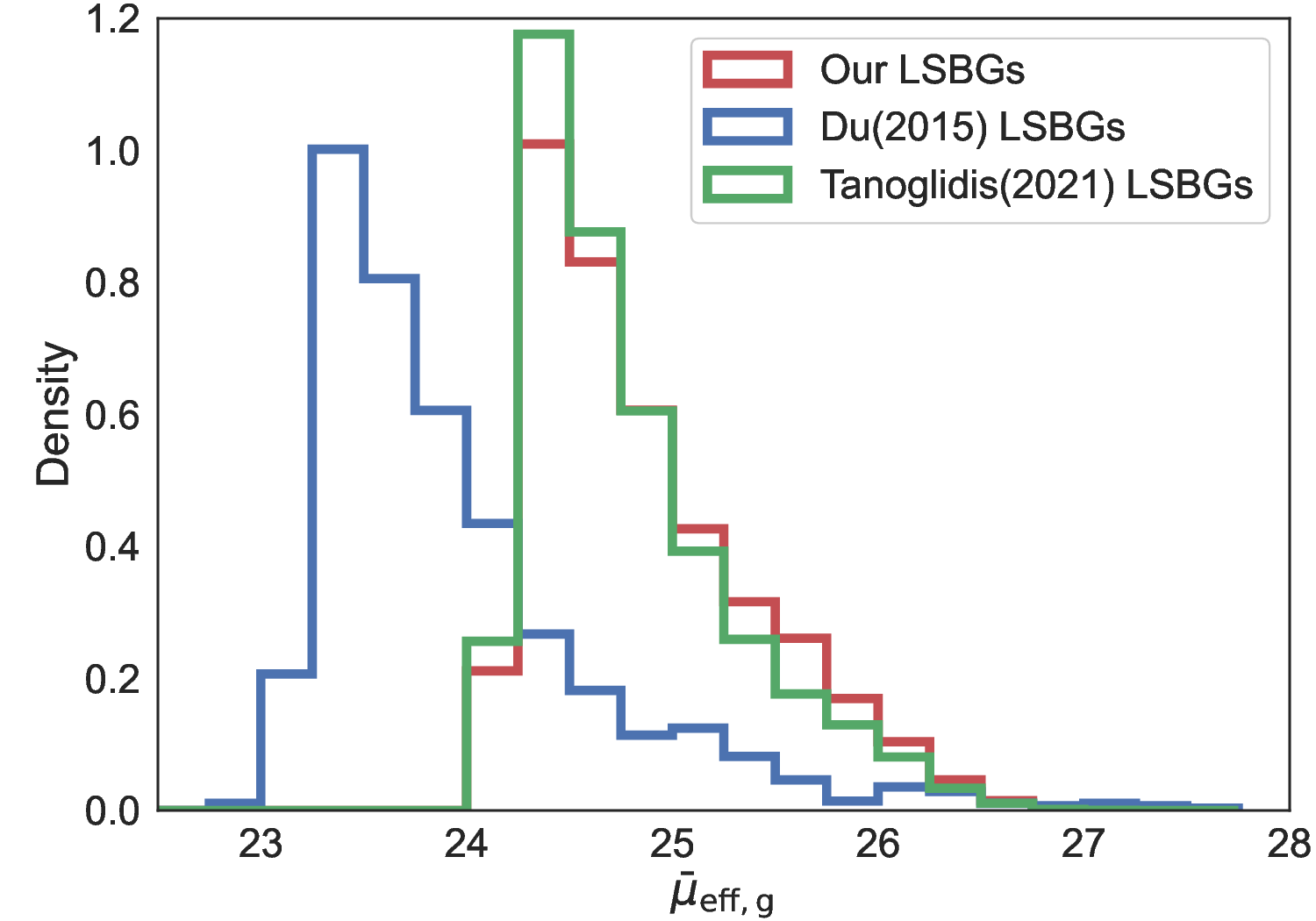}}\hspace{4pt}
    \subcaptionbox{}{\includegraphics[width=.31\linewidth]{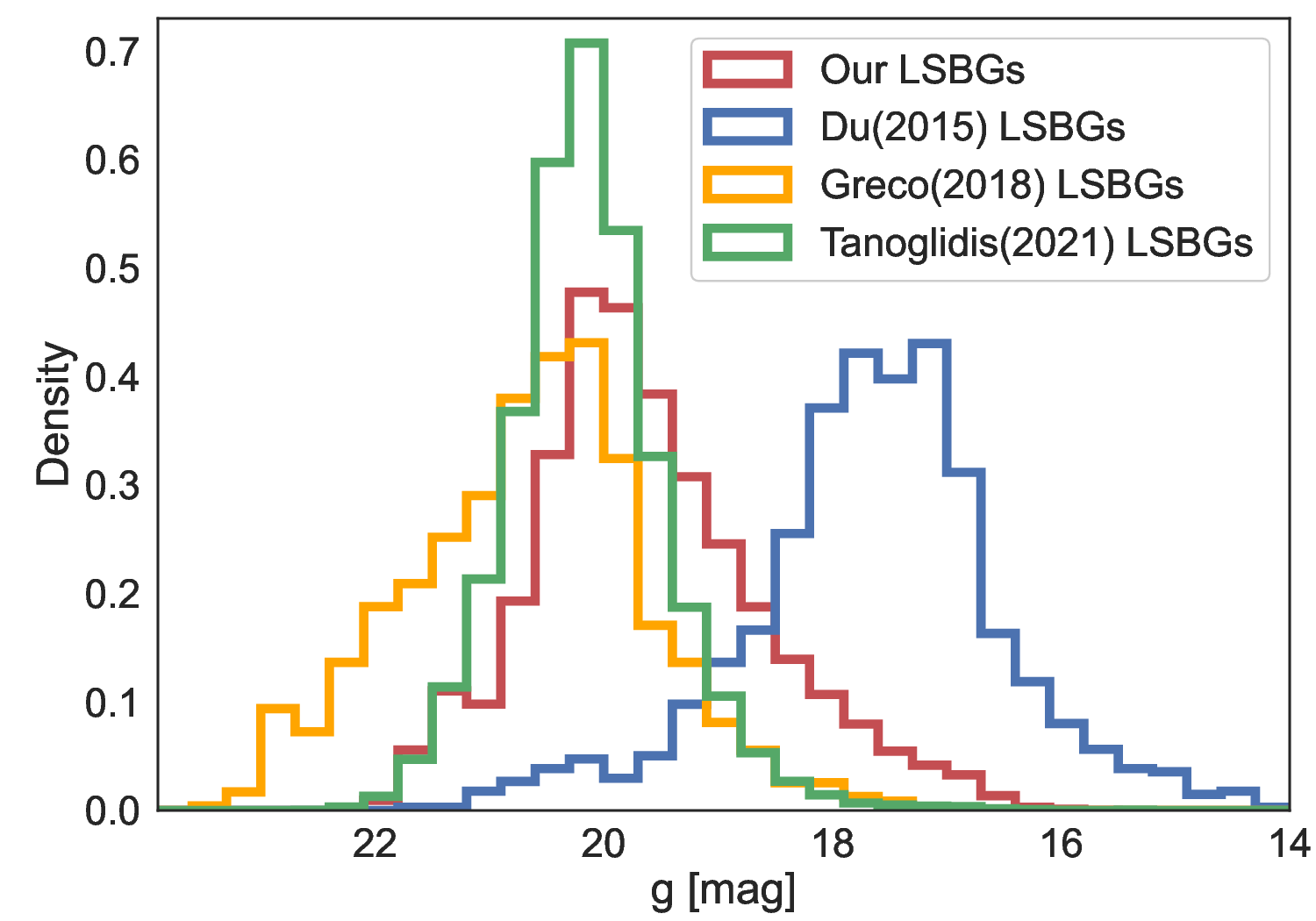}}\hspace{4pt}
    \subcaptionbox{}{\includegraphics[width=.31\linewidth]{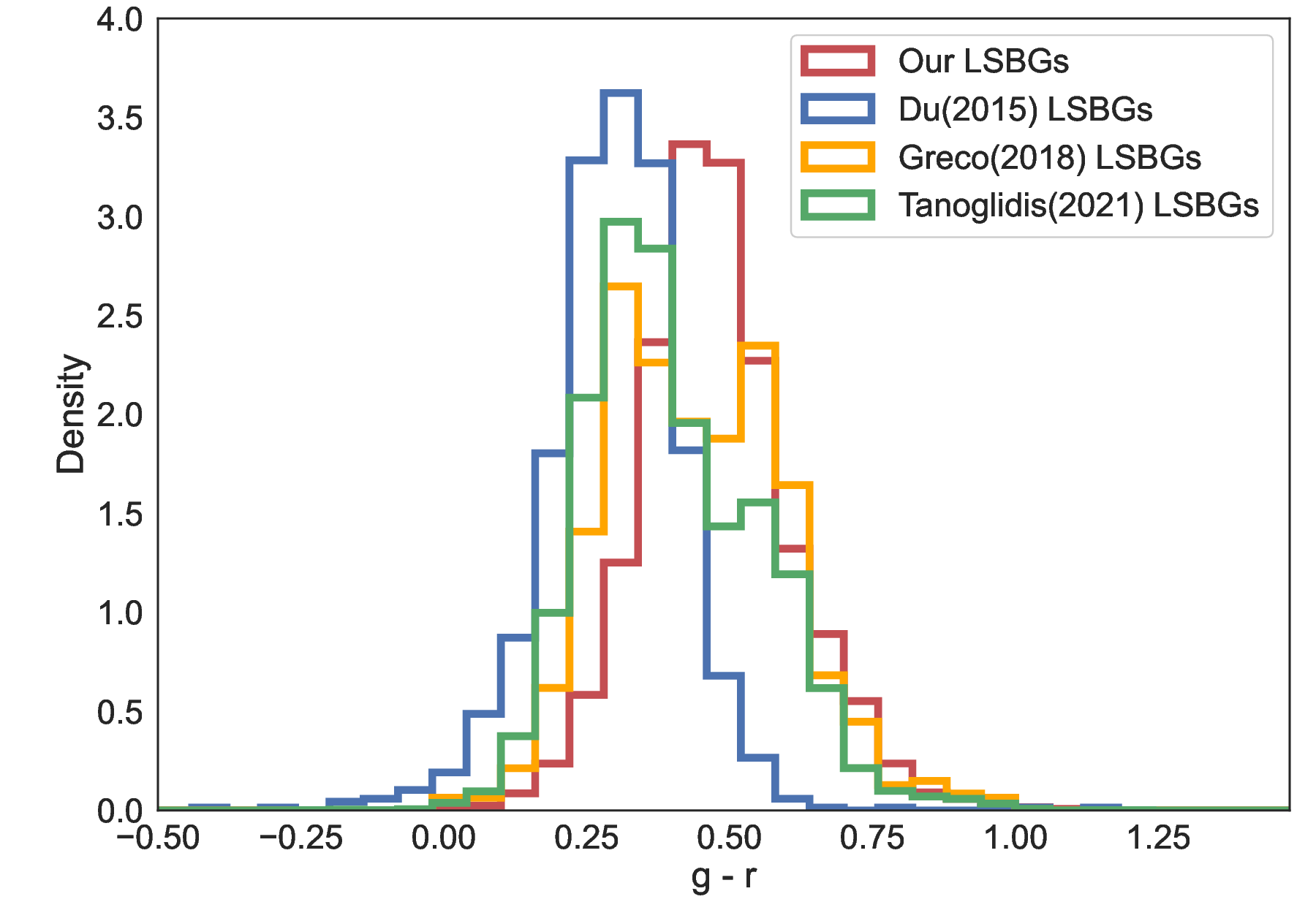}}
    \caption{Comparisons of our sample (red) with the three LSBG samples from \cite{2015AJ....149..199D}(blue), \cite{2018ApJ...857..104G}(orange), and \cite{2021ApJS..252...18T}(green) in terms of the mean surface brightness $\bar{\mu}_{\rm eff,g}$ (left), $g$-band apparent magnitude (middle), and the $g$ - $r$ color distribution (right).}
    \label{Fig11}
\end{figure}

In terms of the surface brightness (Figure \ref{Fig11}(a)), our sample is highly consistent with the T21, ranging from 24.2 $< \bar{\mu}_{\rm eff,g} <$ 28.8 mag arcsec$^{-2}$. The 16$^{th}$, 50$^{th}$, and 84$^{th}$ percentiles of $\bar{\mu}_{\rm eff,g}$ are 24.4, 24.7, and 25.5 mag arcsec$^{-2}$ for our sample and 24.3, 24.7, 25.3 mag arcsec$^{-2}$ for the T21. For the G18 sample, the $\bar{\mu}_{\rm eff,g}$ measurement is not available in its released catalogue, so we are not able to display the G18 sample overplotted in Figure \ref{Fig11}(a) to carry out direct comparisons with the three other samples. However, it is clearly stated in G18 that the $\bar{\mu}_{\rm eff,g}$ distribution of their sample is broad with the 16$^{th}$, 50$^{th}$, and 84$^{th}$ percentiles are $\bar{\mu}_{\rm eff,g}$ = 24.5 (24.8), 24.8 (25.8), and 25.5 (26.8) mag arcsec$^{-2}$ for the blue (red) subsamples. According to such statements, we believe that the G18 sample has quite similar distribution of mean surface brightness to our sample and the T21. In a stark contrast, the Du15 sample has the mean surface brightness distribution with the 16$^{th}$, 50$^{th}$, and 84$^{th}$ percentiles of $\bar{\mu}_{\rm eff,g}$ = 23.4, 23.7, and 24.6 mag arcsec$^{-2}$, which are much brighter than our the three other samples. This is reasonable because the Du15 sample is from the SDSS imaging survey which has a shallower depth than the BASS+MzLS, DES Y3, and HSC-SSP surveys that our sample, T21 and G18 are, repsectively, based on.
This could be furthermore supported by the comparison of magnitude (Figure \ref{Fig11}(b)), where our sample, the T21, and the G18 are systematically at least $\sim$ 2 mag fainter than the Du15 sample in the $g$-band apparent magnitude.  \par

In the aspect of the color (Figure \ref{Fig11}(c)), the 16$^{th}$, 50$^{th}$, and 84$^{th}$ percentiles of $g$ - $r$ are 0.36, 0.47, and 0.60 for our sample,
0.29, 0.43, and 0.60 for the G18 sample, 0.26, 0.38, 0.57 for the T21 sample, and 0.20, 0.30, and 0.41 for the D15.
Apparently, our sample generally agrees with the G18 and the T21 in the $g$ - $r$ distribution, albeit the latter two samples are slightly bluer.  Among the samples for comparison, the sample of Du15 is the bluest because their galaxies are dominated by H{\sc{i}}-rich and blue LSBGs. Additionally, we reported that our sample has a bimodal distribution of the $g$ - $r$ color in Section \ref{subsec:color}, implying two distinct populations of the blue and the red LSBGs, respectively. Actually, such bimodal distributions of the color has also been found in the G18 and the T21 for their own LSBG samples.   
Specifically, the G18 sample shows a clear bimodality in both the $g$ - $r$ and $g$ - $i$ colors,  and is thus divided into two populations of the red and the blue LSBGs using the median $g$ - $i$ = 0.64 as the dividing line. Similarly, the T21 sample also displays bimodal distribution in both the $g$ - $r$ and $g$ - $i$ colors, and is then separated into two subsamples of the blue and the red LSBGs using the intersection of the two Gaussian model profiles at $g$ - $i$ = 0.60 as the threshold. The color distributions of all the three of our sample, the G18, and the T21 demonstrate that LSBGs, similarly to the  galaxies with normal/high surface brightness (normal galaxies), are able to be conventionally divided into two sequences of the blue and the red, with the blue LSBGs dominated by the sprial, disk, or irregular systems in morphology and the red LSBGs by the spheroidal or elliptical morphology.\par

As for the environments, the blue and H{\sc{i}}-rich LSBGs of the Du15 are mostly in voids or to the edge of the filaments of low densities. For the three of our sample, the G18, and the T21, the LSBGs show consistent spatial distributions, with the blue LSBG populations of each sample more uniformly distributed within the sky footprint and the red populations of each sample highly clustered in the spatial area. This implies that the red LSBGs preferentially inhabit in denser environments than the blue LSBGs. 

Furthermore, our sample is consistent with the G18 and the T21 in the ellipticity distribution, with the median around $\epsilon \sim$ 0.3 showing the LSBGs of the three samples are generally round. This is a striking contrast to the almost flat distribution of $\epsilon$ of the normal galaxies between $\epsilon$ = 0 and 0.7.  

These comparisons strongly demonstrate that our sample along with the G18 and the T21 have well extended the SDSS-based LSBG samples to a new regime of much lower surface brightness, fainter apparent magnitude and broad properties in a large scale.

\subsection{Possible evolution from the blue to red LSBGs?} \label{subsec:discussion-evolution}

The optical colors of galaxies indicate their stellar populations and have a strong correlation with the galaxy morphology and environment. In the frame of galaxies with normal or high surface brightnesses, galaxies in the local universe fall into one of two distinct populations in terms of optical colors: a red sequence and a blue cloud \citep{2001AJ....122.1861S,2004ApJ...600..681B,2009ARA&A..47..159B}. Besides the color, the bimodal distributions have also been observed and measured in some other parameters, such as metallicity and star formation rate \citep{2003MNRAS.341...33K,2003MNRAS.341...54K}. The blue cloud is dominated by active, star-forming galaxies while the red sequence is composed of quiescent galaxies. Compared to the blue galaxies which are spiral, disk, or irregular systems in morphology, the red galaxies are ellipticals, spheroidals, lenticulars, and cD galaxies \citep{2009ARA&A..47..159B}. Moreover, red galaxies are more likely to be found in denser environments and more spatially clustered than blue galaxies \citep{2009ARA&A..47..159B,2024arXiv240205788D}. It is proposed that the blue galaxies would evolve onto the red sequence by fading their stellar populations after their star formation was ceased by some quenching mechanisms, such as the natural exhaustion of gas, active galactic nuclei feedback, galaxy harassment, and galaxy mergers, etc.

 Similar to the blue cloud and red sequence in the frame of galaxies with normal surface brightnesses, our LSBGs in this work show a bimodal distribution in the optical color, so they fall into two populations in terms of the $g$ - $r$ color: the blue and red LSBGs. In morphology, the blue LSBGs are disk-like or irregular while the red LSBGs are more bulge-dominated or spheroidal. In addition, the red LSBGs are more spatially clustered than the blue LSBGs. So there might be an possible evolutionary path from the blue LSBGs to the red LSBG, and we would investigate this issue in our future work.

\section{SUMMARY AND CONCLUSIONS}
\label{sect:conclusion}

Based on the released photometric catalogue from the Tractor software and the machine learning model, we selected a sample of 31,825 LSBG candidates with the mean surface brightness  24.2 $< \bar{\mu}_{\rm eff,g} <$ 28.8 mag arcsec$^{-2}$ and the half-light radius 2.5$^{\prime\prime}< r_{\rm eff} <$ 20$^{\prime\prime}$ from the $\sim$ 5500 deg$^{2}$ of the BASS+MzLS survey. The selection criteria are summarized in Table \ref{selectstep}. \par

This sample shows a bimodal distribution in the $g$ - $r$ color, implying two distinct populations of the blue ($g$ - $r <$ 0.60) and red ($g$ - $r >$ 0.60) LSBGs. The blue populations are dominated by spiral, disk or irregular systems while the red ones appear spheroidal or elliptical in morphology, revealing that the colors of LSBGs correlate with morphology. In apparent magnitude and surface brightness, the red LSBGs are slightly fainter than the blue. Both populations have similar distribution of ellipticity, half-light radius (median $r_{\rm eff} \sim$ 4$^{\prime\prime}$), and S$\rm \Acute{e}$rsic index (median $n$ = 1). In terms of ellipticity, the $\epsilon$ for both populations range from 0 to 0.7 with the median $\sim$ 0.3, indicating that the sample galaxies are generally round. This differs from the normal spiral galaxies which show a nearly flat distribution between $\epsilon$ = 0 and 0.7. The half-light radius are within  $\sim2.5^{\prime\prime} - 14^{\prime\prime}$, with a median $r_{\rm eff} \sim$ 4$^{\prime\prime}$. In S$\rm \Acute{e}$rsic index, the blue and the red LSBG populations are both dominated by disk galaxies with $n$ = 1. However, the two populations differ in the spatial distribution, with the blue LSBGs more uniformly distributed across the sky area while the red ones highly clustered. This sample would absolutely be important for further studies on the possible evolutionary link between the two LSBG populations.\par

By comparing our sample with three other samples of LSBGs, it is strongly demonstrated that our sample of LSBG candidates well extends the studies of LSBGs to the regime of lower surface brightness, fainter magnitude, and broader properties than the previously SDSS-based LSBG samples. This sample is definitely an excellent sample for training the deep learning model of higher performance to automatically identify LSBGs from the huge data from more wide and deep imaging surveys in the future.

\begin{acknowledgements}
This work is supported by the National Key R\&D Program of China (grant
No.2022YFA1602901), the Youth Innovation Promotion Association, Chinese Academy of Sciences (No.2020057), the science research
grants from the China Manned Space Project, and the National Natural Science Foundation of China (NSFC; Nos.12090041 and 12090040). Additional support comes from the Strategic Priority Research Program of the Chinese Academy of Sciences
(grant Nos.XDB0550100 and XDB0550102) and the Open
Project Program of the Key Laboratory of Optical Astronomy,
National Astronomical Observatories, Chinese Academy of Sciences.
\end{acknowledgements}


\label{lastpage}

\bibliography{ms2024-0041}{}
\bibliographystyle{raa}



\end{document}